# Neural Network-Based Parametric Model Reduction for Predicting Turbulent Flow for Different Vehicle Geometries

**Kazuto Ando**[1,2,3], **Rahul Bale**[2,3], **Akiyoshi Kuroda**[1], **Makoto Tsubokura**[2,3]

[1] Operations and Computer Technologies Division, RIKEN Center for Computational Science
7-1-26 Minatojima-minami-machi, Chuo-ku, Kobe, Hyogo 650-0047, Japan
[2] Complex Phenomena Unified Simulation Research Team, RIKEN Center for Computational Science,
7-1-26 Minatojima-minami-machi, Chuo-ku, Kobe, Hyogo 650-0047, Japan
[3] Department of Computational Science, Graduate School of System Informatics, Kobe University,
1-1, Rokkodai-cho, Nada-ku, Kobe, 657-8501, Hyogo, Japan

**Keywords:** Parametric model reduction, Neural network, Turbulent flow, Vehicle body

**Abstract.** Numerical simulations in industrial applications often require performing numerous high-precision computations parameterized by specific experimental conditions. For instance, in vehicle body design, aerodynamic simulations are essential for evaluating the aerodynamic characteristics of various proposed body geometries. However, computational resource constraints often become a bottleneck. Therefore, achieving the desired accuracy while minimizing computational cost is crucial. To address this challenge, model reduction methods have been developed to decrease the degrees of freedom by constraining the possible states of a physical system to a lower-dimensional subspace. In particular, reduction techniques that project the system onto a nonlinear subspace using neural networks have been actively studied. Our previous research developed a reduced-order model that integrates neural-network-based model reduction with a time-evolution method, implemented as a distributed parallel training framework to process high-resolution flow field data efficiently. In this study, we extend this reduction approach by incorporating a variational autoencoder to assess its robustness in high-Reynolds-number flows around multiple vehicle bodies with varying geometries. Specifically, we evaluate the reconstruction accuracy of vortex generation across different spatial and temporal scales using a compact latent representation, with a particular focus on the flow behavior near the rear end of the vehicle body.

## 1. Introduction

*Motivation*

Currently, the computational cost required for computational fluid dynamics (CFD) simulations continues to increase. One prominent example is large-scale computations performed on high-end supercomputer systems. Kato et al. [1] conducted large-scale LES (Large Eddy Simulation) on the Supercomputer Fugaku [2,3] using 458,752 MPI processes across 114,688 compute nodes with the flow solver "FrontFlow/blue" to analyze the turbulent flow dynamics around a ship body, utilizing 32 billion finite elements. Similarly, Jansson et al. [4] employed the spectral-element-based direct numerical simulation (DNS) code "Neko" to perform high-precision simulations of Rayleigh-Bénard convection using 16,384 GPUs on LUMI [5].

At the same time, there is an increasing demand for conducting numerous medium-scale simulations, particularly in industrial applications. For instance, multi-objective shape optimization [6] requires running multiple simulations with different shapes to evaluate and optimize aerodynamic performance. However, performing a large number of simulations that fully resolve the smallest scales of the physical phenomena of interest—such as vortices generated by turbulence—entails a substantial computational cost. Kasagi et al. estimate that the spatial scale of streamwise vortices generated around an automobile body ranges from 0.2 mm to 1 mm, while their temporal scale ranges from 100 Hz to 1000 Hz [7], making direct simulations infeasible within a practical computational budget.

*Model reduction methods*

Various methods have been proposed in the CFD community to mitigate the computational burden of such simulations. One widely used approach is projection-based model reduction, which restrict a physical system's states to a specific subspace and represents the system state using a small number of representative variables (hereafter referred to as latent variables). Model reduction methods can be broadly categorized into linear and nonlinear approaches.

A representative method in the former category is proper orthogonal decomposition (POD) [8], which identifies an optimal low-dimensional basis, called modes, to approximate a given dataset—such as time-series flow field data. Originally formulated for extracting coherent structures, POD has been extensively applied to the analysis of turbulent flows and the derivation of low-dimensional dynamical systems via Galerkin projection [9]. For computationally intensive scenarios involving large-scale multidimensional fluid simulation data, the "snapshot POD" approach introduced by Sirovich [10] has become indispensable, enabling efficient mode extraction by solving an eigenvalue problem based on temporal correlations rather than massive spatial dimensions. Building upon this computational foundation, POD-based reduced-order models (ROMs) have been successfully developed to capture complex, transient flow dynamics with high physical fidelity [11]. In contemporary fluid mechanics, POD remains a foundational tool among various data-driven modal analysis techniques [12], and its recent integration with emerging machine learning strategies is rapidly advancing the capability and stability of modeling highly nonlinear flow fields [13].

On the other hand, nonlinear reduction methods, which are well-suited for capturing the nonlinear behavior of flow fields, have also been proposed. A typical approach in this category is the neural-network-based method, which extracts nonlinear modes from time-series instantaneous flow field data. Murata et al. proposed a neural network model called the mode-decomposing convolutional neural network autoencoder (MD-CNN-AE) to reduce the dimensionality of two-dimensional flow data around a circular cylinder [14]. They successfully represented the time evolution of the Kármán vortex street at Re = 100 using only two latent variables without significant reconstruction loss.

In our previous study, MD-CNN-AE was extended to three-dimensional flow fields at a higher Reynolds number (Re = 1,000) [15] using large-scale distributed machine learning on the Supercomputer Fugaku. A three-dimensional flow around a cylinder, simulated using 28 million computational cells, was reduced to 64 latent variables, and the time series of these latent variables were predicted using a long short-term memory (LSTM) network [16]. Our reduction method, implemented via distributed machine learning, demonstrated scalability up to 25,250 computational nodes (1,212,000 cores) on Fugaku, with the convolution routine achieving over 100 PFLOPS in single-precision floating-point arithmetic performance.

Meanwhile, ongoing research focuses on parametric model reduction, which enables reduced-order simulations under different execution settings (e.g., object shape, Reynolds number) than those used during training. One example of this approach is the study by Koo et al. [17], which predicts the time evolution of flow velocity in a reservoir pipe-valve (RPV) system using a POD-based reduced-order model. This model is parameterized with respect to two parameters: the height at the upstream end and the Reynolds number. POD extracts basis functions from instantaneous flow fields generated for nine different parameter combinations using high-precision simulations. Finally, the accuracy of the reduced-order model was evaluated for parameter values not included in these nine training cases. A detailed classification of parametric model reduction methods is provided by Benner et al. [18], where techniques such as rational interpolation, balanced truncation, and POD are introduced, along with application examples including thermal modeling of electric motors, optimization of batch chromatography, and control of rod movement in nuclear reactor cores.

Also, efforts have been made to optimize automobile body shapes in the latent space. Tran et al. [19] obtained a low-dimensional representation of vehicle geometries using an autoencoder and, by learning the mapping between this representation and aerodynamic characteristics obtained from LES simulations, enabled the derivation of vehicle body shapes that meet the required aerodynamic performance through optimization in the latent space.

These studies primarily focus either on flows around simple geometries, such as bluff bodies, at Reynolds numbers on the order of a few hundred, or on the prediction of time-averaged flow fields rather than instantaneous flow fields. In contrast, from the perspective of industrial applications of simulation, more practical problems—namely, high-Reynolds-number flows around complex geometries that require detailed spatiotemporal resolution—have not been addressed using neural network-based reduced-order models. This is largely due to the lack of technologies for highly parallel distributed training and the computational resources necessary to realize them.
In this study, we present a neural network-based reduced-order model designed for highly parallel distributed training. By incorporating a variational autoencoder (VAE), the proposed approach improves robustness to variations in object geometry. Furthermore, using multiple real-world automobile geometries, we demonstrate its applicability to a highly practical problem from the perspective of industrial simulation–namely, the prediction of high-Reynolds number flows around previously unseen complex geometries.

*Related works*
Hasegawa et al. [20] proposed a neural-network-based model reduction technique that combines a convolutional neural network autoencoder with a long short-term memory (LSTM) network. They evaluated the robustness of their model for two-dimensional flow around various randomly generated bluff body shapes. As a result, flow fields for approximately 20 bluff body shapes were successfully reproduced using a neural network trained on 80 shapes.

Hasegawa et al. [21] further assessed the robustness of their method for two-dimensional circular cylinder flow at varying Reynolds numbers. In this study, Reynolds number information was fed into the LSTM network, allowing the model to successfully predict the flow field around a circular cylinder at Reynolds numbers not included in the training data.

Higashida et al. [22] investigated the robustness of our previously developed model reduction method [15] for two-dimensional flow around two distant square cylinders with varying inter-cylinder distances. A comprehensive comparison with POD across different inter-cylinder distances demonstrated that the proposed method outperforms POD in reconstructing flow fields at untrained distances.

*Contributions*
This study proposes a neural-network-based parametric model reduction method tailored for three-dimensional high-Reynolds-number turbulent flow. This paper makes the following two contributions.

First, a variational autoencoder (VAE) is introduced for dimensionality reduction. The enhancement in flow reconstruction accuracy obtained by incorporating a VAE into our previously proposed network architecture has been reported in a domestic conference [23]. A brief overview is provided below.

The reconstructed flow fields obtained using conventional methods and the proposed approach are presented in Figure 1. In all cases, the flow fields are reconstructed from a reduced representation consisting of 64 variables using Proper Orthogonal Decomposition (POD) and neural-network-based methods (MD-CNN-AE and MD-CNN-AE+VAE), enabling a direct and fair comparison of reconstruction capability under the same dimensionality constraint.

Compared with the high-precision simulation results (full-order model), POD fails to reproduce the complex wake structures behind the cylinder, particularly in regions characterized by strong nonlinear interactions. In contrast, the neural-network-based approach (MD-CNN-AE) captures significantly richer flow features, demonstrating its ability to represent nonlinear flow dynamics beyond the limitations of linear subspace methods.

Furthermore, the proposed method (MD-CNN-AE+VAE) achieves a further improvement in reconstruction fidelity. It more accurately recovers the spatial complexity of the wake structures and shows closer agreement with the full-order model across all observed regions. Notably, the spanwise flow components (vertical direction in the figures, bottom row), which are difficult to reconstruct using conventional approaches, are also successfully reproduced. These results indicate that the incorporation of the variational autoencoder enhances the robustness and expressiveness of the reduced-order model, particularly for capturing high-dimensional and nonlinear flow features.

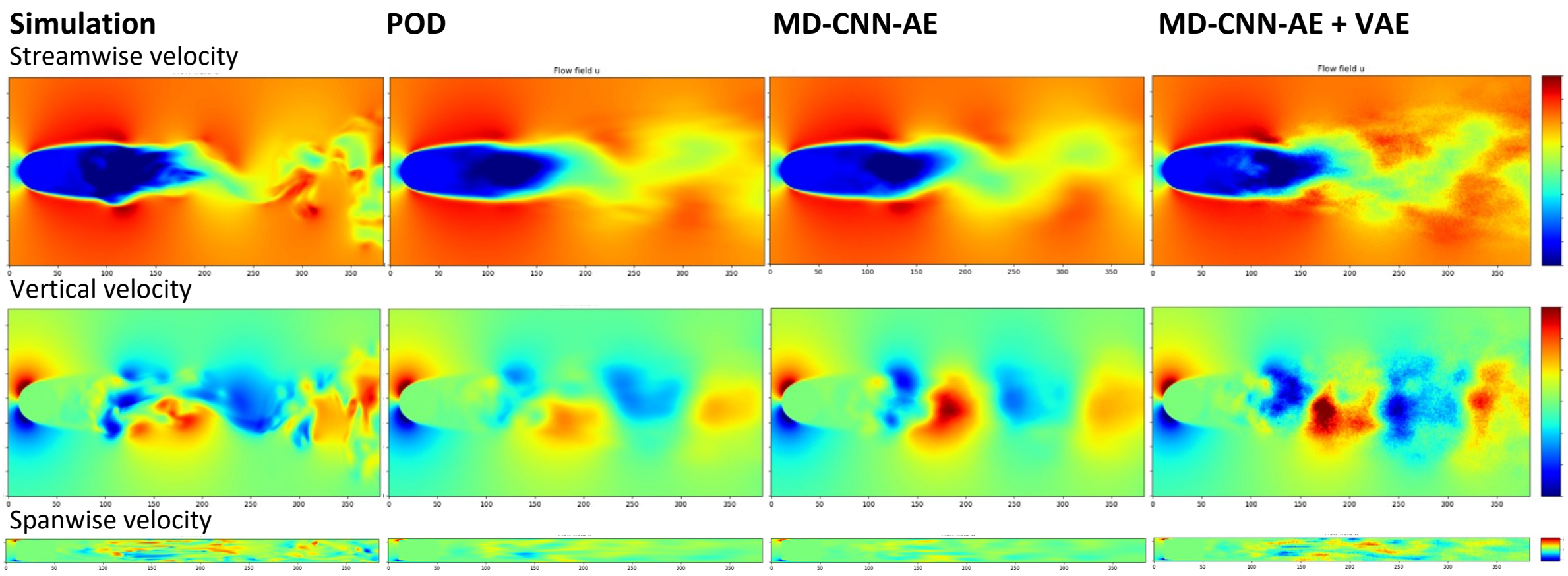

**Figure 1.** Comparison of reconstructed flow fields generated by conventional and proposed methods for the reduced-order representation of the high-fidelity flow field. The rows correspond to the streamwise, vertical, and spanwise velocity components, while the columns compare the reference simulation results with reconstructions obtained using POD, MD-CNN-AE , and the proposed MD-CNN-AE + VAE method.

Second, we demonstrated the applicability of the proposed method using eleven different realistic vehicle models. The proposed method predicts the time evolution of turbulent flow (Re ~ $8.4 \times 10^6$) in the rear-end region of a vehicle body with varying shapes. To the best of our knowledge, applying a neural-network-based reduced-order model to predict three-dimensional turbulent flows around complex geometries at such high Reynolds numbers, particularly for unseen configurations, has not been reported in previous studies. This capability is enabled for the first time by our network architecture and its distributed parallel training algorithm.

*Organization of the Paper*

The paper is structured as follows: Section 2 introduces the proposed model reduction method, including training data generation, network architecture, and parallelization strategies. Section 3 presents the evaluation results of the proposed model reduction method for turbulent flow with varying vehicle body shapes. Finally, Section 4 concludes the paper with a summary.

## 2. Methods

This section describes the method used to generate training data for this study, specifically the target flow field data for model reduction. It also introduces our neural-network-based parametric nonlinear reduction method, providing a detailed explanation of the neural network architecture and distributed parallel training approach.

### *2.1. Training data*

This section explains how the training data used in this study was generated. High-precision three-dimensional turbulent flow simulations were conducted using the simulation framework "CUBE" [24], developed by the Complex Phenomena Research Team at the RIKEN Center for Computational Science. CUBE is a unified simulation framework based on the Building Cube Method (BCM) [25,26] and the Immersed Boundary Method (IBM) [27,28].

*Governing equations*
The governing equations for the flow simulations conducted in this study are defined as follows. The incompressible Navier-Stokes equations are given by

$$\rho\left(\frac{\partial \boldsymbol{u}}{\partial t}+(\boldsymbol{u}\cdot\nabla)\boldsymbol{u}\right)=-\nabla p+\mu\nabla^2\boldsymbol{u}+\boldsymbol{f}, \tag{1}$$

$$\rho\nabla\cdot\boldsymbol{u}=0, \tag{2}$$

where $\boldsymbol{u}$ represents the flow velocity vector, $p$ is the pressure, $\rho$ is the density, $\mu$ is the viscocity, and $\boldsymbol{f}$ is the force vector exerted by the immersed geometries in the fluid (the vehicle body in this study).
The above equations are solved using a finite volume method on a hierarchical structured grid. No explicit turbulence model is employed; instead, an implicit large-eddy simulation (ILES) [29] approach is adopted, in which the effects of subgrid-scale vortices are implicitly accounted for through numerical dissipation.

*Problem settings*
This section describes the problem settings in this study. The computational domain extends over a large area surrounding the vehicle body, with a constant inflow entering in the positive X-direction (Figure 2(a)).

In contrast, the region used for training the machine learning model is restricted to the area behind the vehicle body (shown as the gray area in Figure 2(a)). This region is selected both due to computational resource constraints, which necessitate restricting the domain, and because, as reported in [30], the flow in this region is a dominant contributor to drag and is characterized by concentrated vortex shedding and recirculation. The computational grid becomes increasingly finer as it approaches the vehicle body's surface, and within the training domain, the grid spacing is uniform and at its finest resolution (Figure 2(b)).

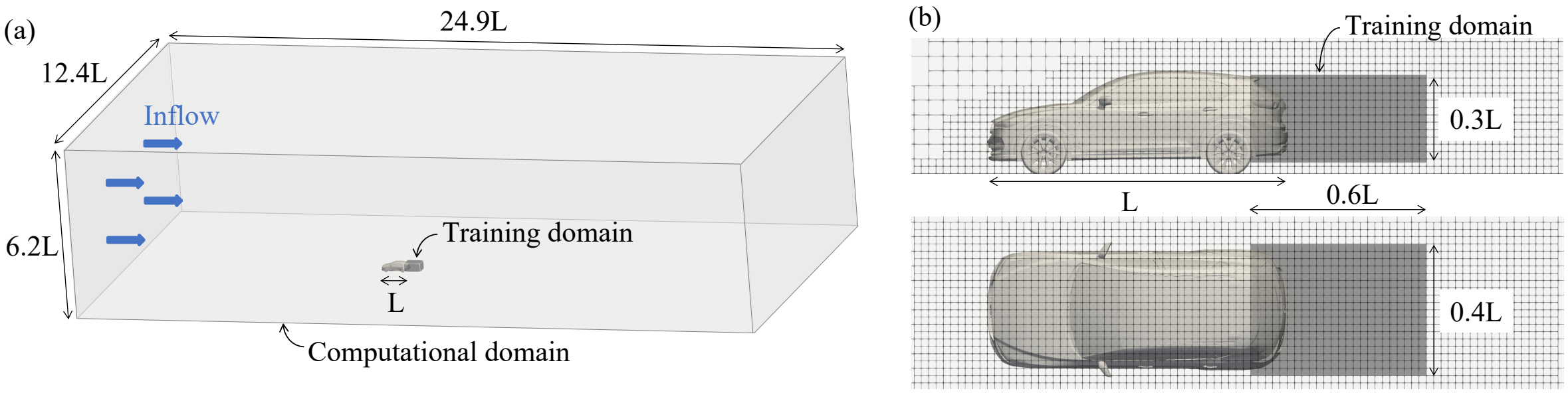

**Figure 2.** (a) Computational domain showing the inflow direction and overall geometry. (b) Training domain with uniformly refined grid spacing, highlighting the region used for model training.

In this study, three-dimensional flow velocity fields were obtained from turbulent simulations performed under the same execution conditions (see Table 1) for 11 vehicle body models provided by Professor Takuji Nakashima from Hiroshima University (Figure 3(a)).

These models were generated through principal component analysis (PCA) based on data from 124 commercially available SUVs, each characterized by 22 side-view design variables, including roof length, hood length, and nose height (Figure 3(a)). For each vehicle geometry, a vector representing the set of design variables is defined, and these vectors are concatenated column-wise to construct a matrix. PCA is then applied to this matrix to obtain eigenvectors and eigenvalues (principal component scores) that characterize the vehicle geometries. The first principal component score was varied from -1.5 to 1.5 in increments of 0.3, resulting in 11 distinct models.

These variations in shape lead to significant differences in aerodynamic performance. For instance, the maximum deformation in the X-direction at the rearmost position of the vehicle body is approximately 12 cm (Figure 3(b)).

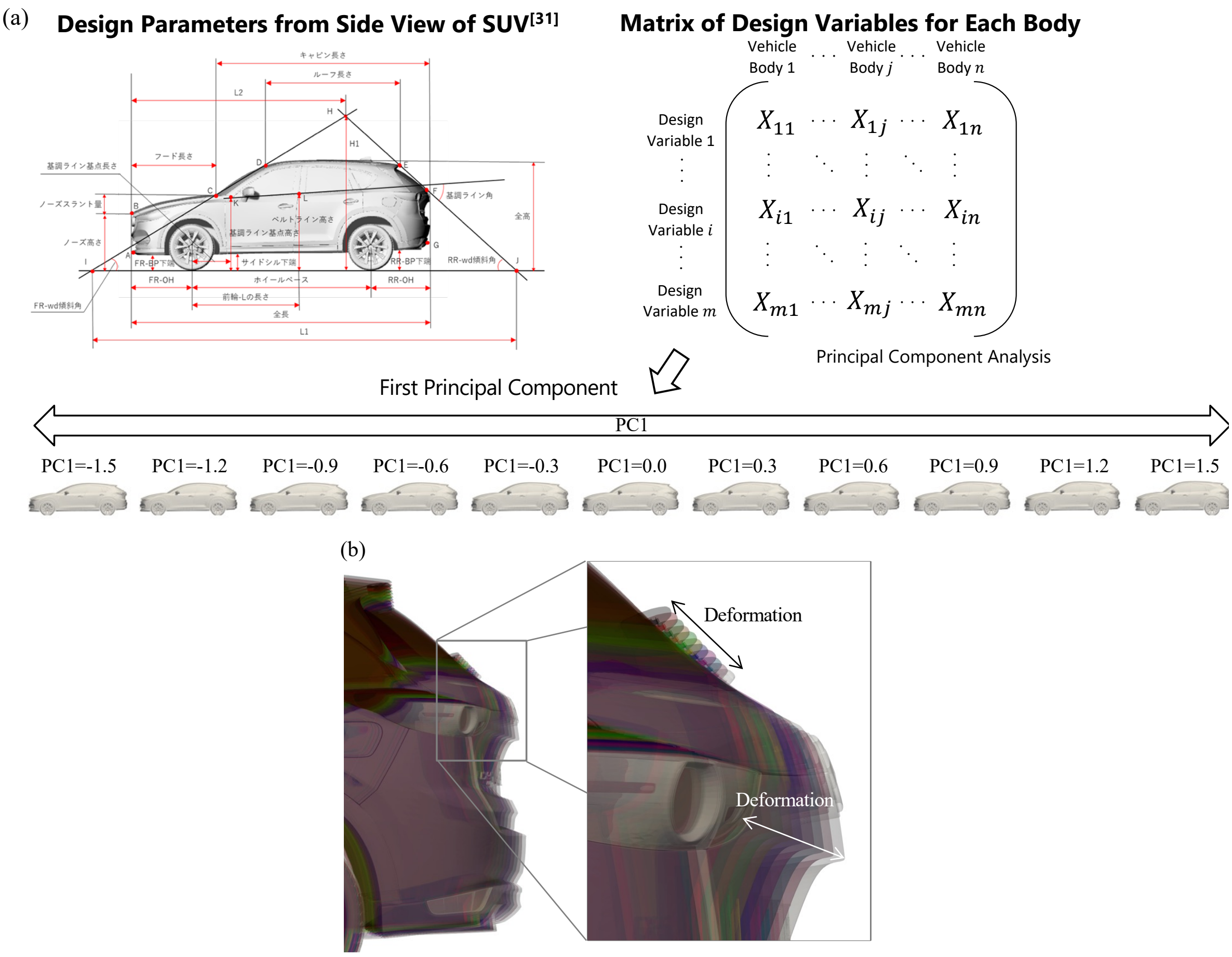

**Figure 3.** (a) Vehicle body geometries generated using principal component analysis (PCA), where each model corresponds to a different principal component (PC1) coefficient. (b) Detailed view of shape variations at the rearmost region of the vehicle body, illustrating the deformation associated with changes in the PCA coefficient.

Table 1 provides a detailed description of the simulation settings used in this study. The computational domain is discretized into 200 million cells, with a Reynolds number of approximately $8.4 \times 10^6$. The sampling frequency for the training data is set to 0.5 ms to capture high-frequency fluctuations in the wake region.

**Table 1.** Simulation parameters and conditions.

| | |
|---|---|
| Flow solver type | Incompressible flow solver |
| Computational domain size | 114 m × 57 m × 28 m |
| Vehicle model size | 4.6 m × 2.0 m × 1.5 m |
| Number of cubes | 48,980 |
| Cells per cube | 16 × 16 × 16 = 4,096 |
| Total number of cells | 48,980 × 4,096 = 200,622,080 |
| Minimum cell size | 7.0 mm |
| Time step size | $2.5 \times 10^{-5}$ s |
| Sampling frequency | 0.5 ms (per 20 steps) |
| Integration time | 1.75 s (70,000 steps) |
| Inflow velocity | 27.78 m/s (100 km/h) |
| Reynolds number | $\sim 8.4 \times 10^6$ |
| Time integration scheme | Crank-Nicolson method |
| Pressure Poisson solver | V-cycle multigrid |
| Viscous term discretization | Second-order central difference |
| Convection term discretization | QUICK scheme |

## *2.1. Training strategy*

This section provides details on network training, including the network architecture, loss function, and distributed parallel training method.

*Network architecture*

The schematic of the entire network architecture used in this study is shown in Figure 4.

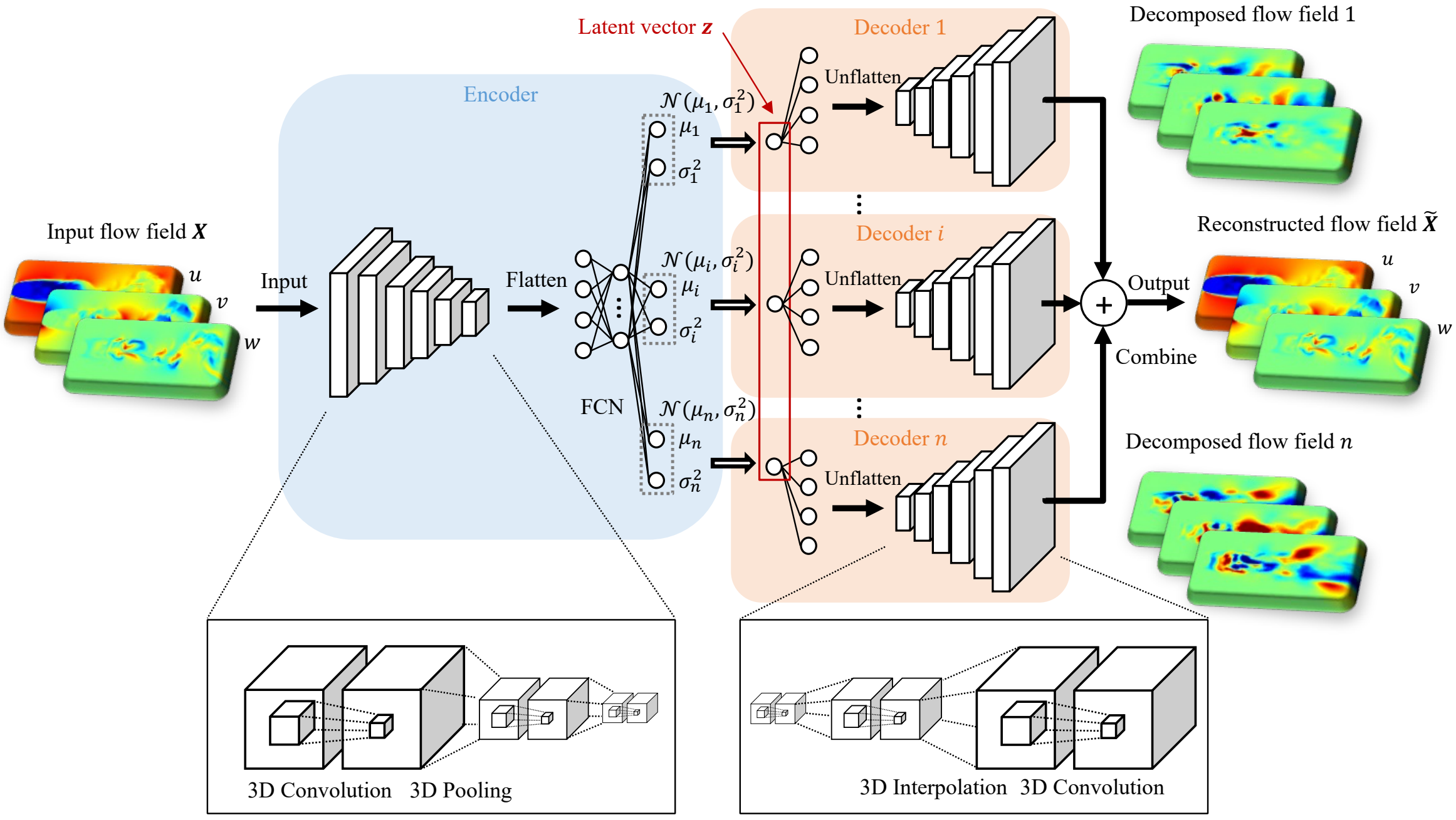

**Figure 4.** Overview of the proposed neural network architecture, consisting of an encoder that extracts latent representations from the input data and a branched decoder structure that reconstructs multiple output fields from each latent vector element.

The first half of the network, called the encoder, reduces the high-dimensional flow field data and outputs a low-dimensional representation called the latent vector. The encoder consists of multiple layers of blocks composed of three-dimensional convolutional layers and pooling layers. To stabilize training, layer normalization is applied to the output of each layer.

In contrast, the latter half of the network, called the decoder, is branched, with each branch having its own parameters (where the branched decoder has the same structure as that of MD-CNN-AE). Each branch expands the dimension from a single element of the latent vector to a decomposed flow field, which has the same dimensions as the original high-dimensional flow field data. The decoder consists of multiple layers of blocks composed of convolutional layers and interpolation layers. Finally, the decomposed flow fields are combined, producing the reconstructed flow field. Training is performed to minimize the error between the original and reconstructed flow fields. Once training has sufficiently progressed, the network can effectively map high-dimensional flow field data to a low-dimensional representation.

This procedure corresponds to the decomposition of the flow field into multiple nonlinear modes, with mode information preserved as network parameters. The number of decompositions in the decoder is a user-defined hyperparameter, which is equal to the dimensionality of the latent vector. The smaller this value, the higher the reduction (compression) efficiency. However, there is a trade-off: reducing the latent dimension decreases reconstruction accuracy. It should be noted that a large number of branches (modes) is required to efficiently reduce highly nonlinear flows, such as high-Reynolds-number turbulent flows. The detailed network structure is provided in Table A.1 and A.2 in Appendix A.

*Formulation of the Problem to Be Solved by the Neural Network*
The loss function is formulated as follows:

$$L\left(\boldsymbol{X}(t), \widetilde{\boldsymbol{X}}(t), \boldsymbol{\phi}, \{\boldsymbol{\theta}_i\}_{i=1}^{n}\right) = \frac{\left\|\boldsymbol{X}(t) - \widetilde{\boldsymbol{X}}(t)\right\|_2^2}{N} + D_{KL}\left(q_\phi\left(\boldsymbol{z}(t)\middle|\boldsymbol{X}(t)\right)\middle\|p\left(\boldsymbol{z}(t)\right)\right), \tag{3}$$

where $\boldsymbol{X}(t)$ is the high-precision simulation result of the flow field (i.e., the input data of the network) at time $t$, while $\widetilde{\boldsymbol{X}}(t)$ is the reconstructed flow field (i.e., the output data of the network) at time $t$.

$D_{KL}$ represents the Kullback–Leibler (KL) divergence,

$$D_{KL}\left(q_\phi\left(\boldsymbol{z}(t)\middle|\boldsymbol{X}(t)\right)\middle\|p\left(\boldsymbol{z}(t)\right)\right) = \int q_\phi\left(\boldsymbol{z}(t)\middle|\boldsymbol{X}(t)\right) \log \frac{q_\phi\left(\boldsymbol{z}(t)\middle|\boldsymbol{X}(t)\right)}{p\left(\boldsymbol{z}(t)\right)} d\boldsymbol{z}, \tag{4}$$

where $q_\phi\left(\boldsymbol{z}(t)\middle|\boldsymbol{X}(t)\right)$ is the posterior distribution of the latent vector $\boldsymbol{z}(t)$ given the input flow field $\boldsymbol{X}(t)$.

$p\left(\boldsymbol{z}(t)\right)$ is the standard normal distribution, representing the prior probability distribution of the latent vector $\boldsymbol{z}(t)$:

$$p\left(\boldsymbol{z}(t)\right) = \mathcal{N}(\boldsymbol{z}(t); \boldsymbol{0}, \mathbf{I}). \tag{5}$$

The optimization problem solved during training is formulated as follows:

$$\boldsymbol{\phi}, \{\boldsymbol{\theta}_i\}_{i=1}^{n} = \operatorname{argmin}_{\boldsymbol{\phi}, \{\boldsymbol{\theta}_i\}_{i=1}^{n}} \sum_{t_{min}}^{t_{max}} L\left(\boldsymbol{X}(t), \widetilde{\boldsymbol{X}}(t), \boldsymbol{\phi}, \{\boldsymbol{\theta}_i\}_{i=1}^{n}\right), \tag{6}$$

where $\boldsymbol{\phi}$ represents the network parameters of the encoder, $\boldsymbol{\theta}_i$ is the network parameters of the decoder corresponding to the $i$-th mode, and $n$ is the number of modes (i.e., the number of decoders or the number of latent variables). $t_{min}$ and $t_{max}$ denote the start and end points of the training time interval, respectively.

The reconstructed flow field $\widetilde{\boldsymbol{X}}(t)$ can be expressed as follows:

$$\widetilde{\boldsymbol{X}}(t) = \sum_{i=1}^{n} \boldsymbol{\mathcal{F}}_{dec}(z_i(t); \boldsymbol{\theta}_i), \tag{7}$$

where $\boldsymbol{\mathcal{F}}_{dec}$ represents the decoder neural network, and $z_i(t)$ is the latent variable corresponding to the $i$-th mode at time $t$.

$\mathcal{N}(\boldsymbol{z}(t); \boldsymbol{\mu}(t), \boldsymbol{\sigma}(t)^2)$ is the normal distribution with mean $\boldsymbol{\mu}(t)$ and variance $\boldsymbol{\sigma}(t)^2$, which represents the probability distribution of the latent vector $\boldsymbol{z}(t)$:

$$\boldsymbol{z}(t) \sim \mathcal{N}(\boldsymbol{z}(t); \boldsymbol{\mu}(t), \boldsymbol{\sigma}(t)^2). \tag{8}$$

$\boldsymbol{\mu}(t)$ and $\boldsymbol{\sigma}(t)^2$ are obtained as follows:

$$\boldsymbol{\mu}(t), \boldsymbol{\sigma}(t)^2 = \boldsymbol{\mathcal{F}}_{enc}(\boldsymbol{X}(t); \boldsymbol{\phi}), \tag{9}$$

where $\boldsymbol{\mathcal{F}}_{enc}$ represents the encoder neural network.

During the inference phase, the input flow field data is reconstructed using the trained network parameters $\boldsymbol{\phi}$ and $\{\boldsymbol{\theta}_i\}_{i=1}^{n}$ as follows:

$$\widetilde{\boldsymbol{X}}(t) = \sum_{i=1}^{n} \boldsymbol{\mathcal{F}}_{dec}(\mu_i(t); \boldsymbol{\theta}_i), \tag{10}$$

$$\boldsymbol{\mu}(t), \boldsymbol{\sigma}(t)^2 = \boldsymbol{\mathcal{F}}_{enc}(\boldsymbol{X}(t); \boldsymbol{\phi}). \tag{11}$$

Note: In the inference phase, the latent variable used as input to the decoder is the mean of the normal distribution representing the probability of the latent variables. In other words, during training, the $i$-th component of the random variable $\boldsymbol{z}(t)$ sampled from the normal distribution defined in Eq. (8) is provided as an input to Eq. (7), whereas during inference, the $i$-th component of mean value $\boldsymbol{\mu}(t)$ of the distribution is used instead. This means that the decoder operates deterministically, rather than using probabilistic sampling.

*Distributed parallel training*

All the experiments in this study were conducted on the Supercomputer Fugaku (hereafter, Fugaku). Fugaku is equipped with a single Fujitsu A64FX CPU per node. The A64FX processor has 48 computational cores, each achieving 128 GFLOPS in single-precision floating-point arithmetic performance. Consequently, each node delivers 6.1 TFLOPS of single-precision performance.

For comparison, a high-end GPU, the NVIDIA H100 NVL [32], attains 60 TFLOPS for single precision (excluding tensor-core operations), which is approximately 10 times higher than the performance of a single A64FX processor. To close this gap, it is necessary to utilize a large number of nodes on Fugaku, requiring an efficient distributed (inter-node) parallel training mechanism.

To address this issue, in our previous work, a distributed parallel training framework was implemented using hybrid parallelism, which combines data parallelism and network architecture-specific model parallelism [15]. This framework has demonstrated scalability, allowing training execution to scale up to tens of thousands of nodes (millions of cores).

Table 2 presents the execution settings for distributed training. Notably, this study aims to reconstruct a flow field represented by 20 million variables using only 32 latent variables. Two experiments are conducted for comparison:

- Experiment 1: Training on flow fields around six vehicle body shapes.
- Experiment 2: Training on flow fields around two vehicle body shapes.

The sampling interval of the instantaneous fields used as training data is identical to the output interval of the simulation. Specifically, a value of 0.5 ms, denoted as “Sampling frequency” in Table 1, is adopted. As mentioned above, this choice is intended to resolve vortices at 100 Hz over 20-time steps.

A detailed explanation of the hyperparameters used for neural network training is provided in Table A.3 in Appendix A.

**Table 2.** Execution settings for distributed training.

| Target physical variables for training | Three-dimensional flow velocity ($u$, $v$, $w$) |
|---|---|
| Target region | 0.6L × 0.4L × 0.3L |
| Grid resolution | 384 × 288 × 192 = 21,233,664 |
| Number of latent variables (i.e., number of decomposition modes) | 32 |
| Grid spacing | 7.0 mm (uniform resolution) |
| Sampling interval | 0.5 ms |
| Shapes used for inference | Experiment 1: PC1 = -1.5, -0.9, -0.3, 0.3, 0.9, 1.5<br>Experiment 2: PC1 = -1.5, 1.5 |
| Shapes for inference | PC1 = -1.2, -0.6, 0.0, 0.6, 1.2 |
| Training samples per shape | 492 |
| Total training samples | Experiment 1: 492 × 6 = 2,952<br>Experiment 2: 492 × 2 = 984 |
| Total training iterations | 12,000 (2,000 epochs) |
| Number of snapshots generated during inference | 500 |
| Integration time for inference | 0.25 s |

Figure 5 illustrates which of the 11 vehicle body shapes were used for training or testing in each experiment.

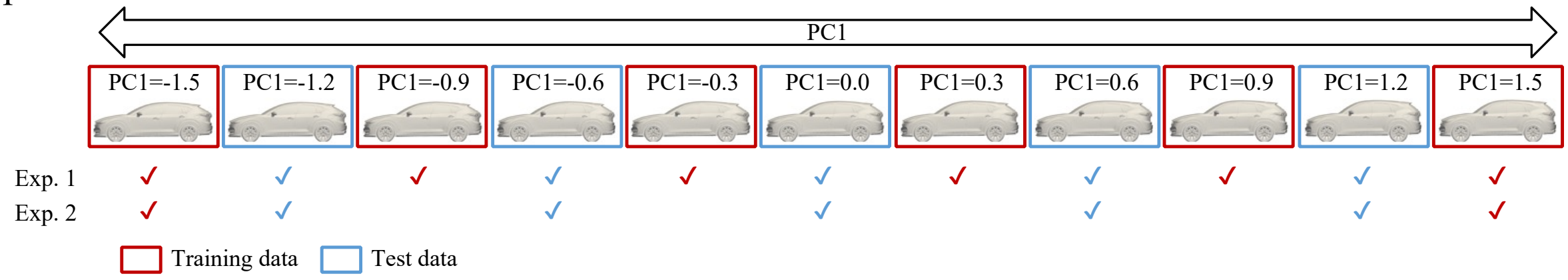

**Figure 5.** Vehicle body shapes used for training and testing, parameterized by the principal component (PC1) coefficient. Shapes highlighted in red and blue indicate the subsets assigned to the training and testing datasets, respectively.

The computational resources required during training are summarized in Table 3. These experiments demanded substantial computational power: Experiment 1 consumed 1 million node hours, while Experiment 2 required 340,000 node hours on Fugaku.

**Table 3.** Computational resource requirements for training.

| | |
|---|---|
| Number of compute nodes used | Experiment 1: 8,118 nodes (389,664 cores)<br>Experiment 2: 2,706 nodes (129,888 cores) |
| Theoretical peak single-precision performance | Experiment 1: 6.1 TFLOPS × 8,118 nodes = 49.8 PFLOPS<br>Experiment 2: 6.1 TFLOPS × 2,706 nodes = 16.6 PFLOPS |
| Training time per epoch | Experiment 1: 48 s<br>Experiment 2: 46 s |
| Total training iterations | 60,000 iterations (10,000 epochs) |
| Total training time | Experiment 1: (48 / 3600) hour × 10,000 epochs = 133 hours<br>Experiment 2: (46 / 3600) hour × 10,000 epochs = 127 hours |
| Estimated total computational cost | Experiment 1: 8,118 nodes × 133 hours = 1.0 × $10^6$ node-hours (corresponds to 6.6 ExaFLOP)<br>Experiment 2: 2,706 nodes × 127 hours = 3.4 × $10^5$ node-hours (corresponds to 2.2 ExaFLOP) |

**3. Results and discussion**

This chapter presents the evaluation results of the proposed model reduction method for the two experiments. The first experiment, referred to as "Experiment 1," evaluates model reduction using a network trained on flow fields around six vehicle body shapes. The second experiment, referred to as "Experiment 2," evaluates model reduction using a network trained on flow fields around two vehicle body shapes.

*3.1. Training Progress and Convergence Status*

Figure 6 illustrates the decay of the normalized training error as the number of training iterations progresses in Experiment 1 and Experiment 2. The training error decreases monotonically with each iteration; however, at the current stage of training (12,000 iterations), the decrease in training error remains incomplete and has not yet saturated in either experiment. Therefore, the accuracy of the model reduction presented in this chapter can be further improved by continuing the training. At 12,000 iterations, more than 99% of the total loss is attributed to the reconstruction loss. This indicates that, at this stage, the reconstruction loss aimed at improving flow-field fidelity is overwhelmingly dominant over the KL divergence term, which regularizes the latent distribution toward a standard normal distribution. As training progresses, however, the reconstruction loss is expected to decrease by orders of magnitude, at which point the influence of the KL divergence constraint is expected to become more significant.

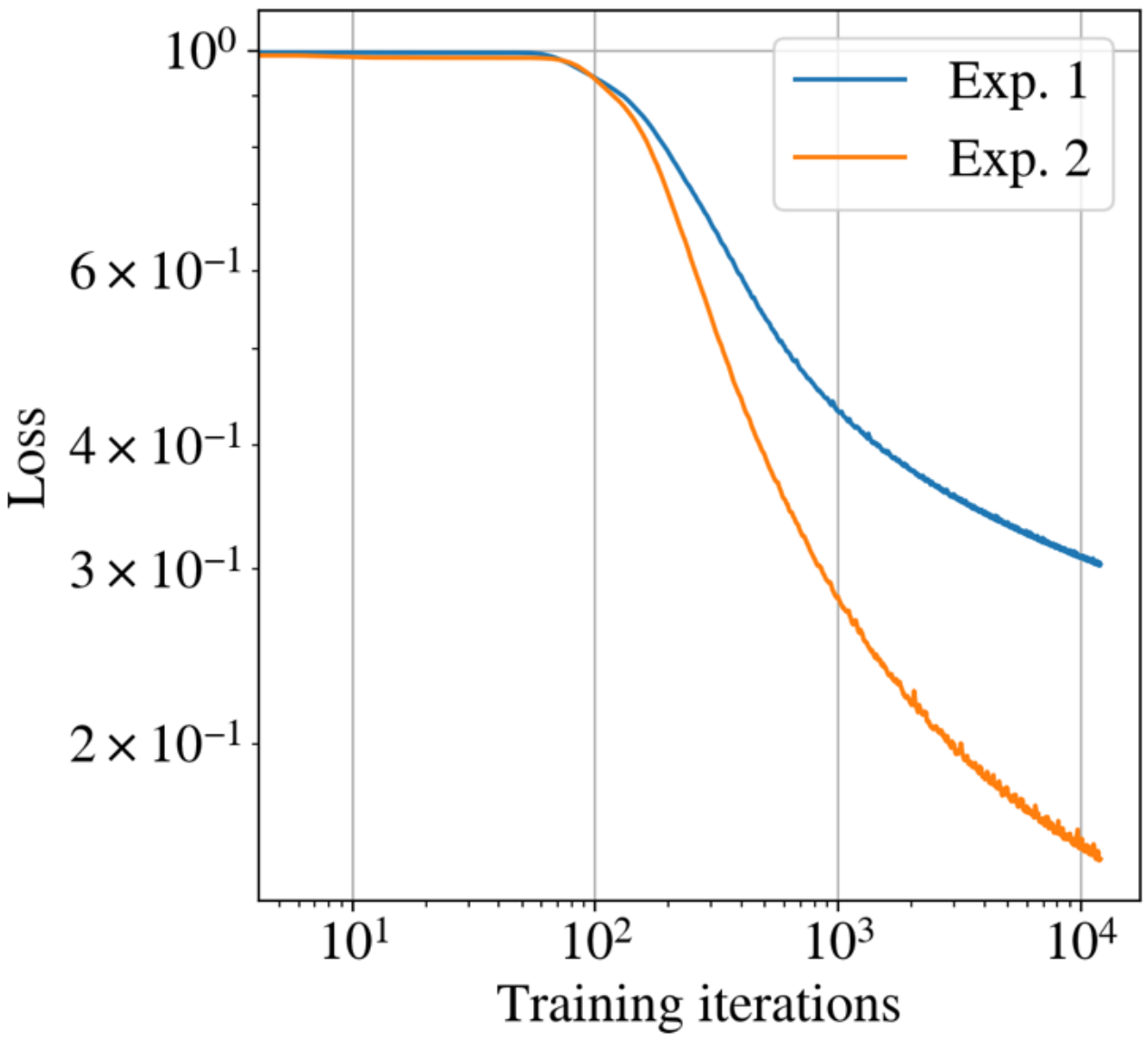

**Figure 6.** Comparison of loss convergence over training iterations for Exp. 1 (blue) and Exp. 2 (orange). While both experiments exhibit similar loss values in the early stage, their behaviors diverge after approximately $10^2$ iterations.

*3.2. Extracted modes*

Figure 7 illustrates the time variation of selected latent vector elements corresponding to the flow field around the PC1=0.0 shape as an example. This value corresponds to $z_i(t)$ in Equation (7).

Comparing the results of Experiment 1 and Experiment 2, some elements, such as $z_1$ exhibit correlations between the values from both experiments. In contrast, certain elements, such as $z_4$, appear to have no correlation. This is likely because the spatial scale of flow fluctuations associated with a specific mode number depends on runtime conditions. For the graphs of all latent variables, see Figure B.1 in Appendix B.

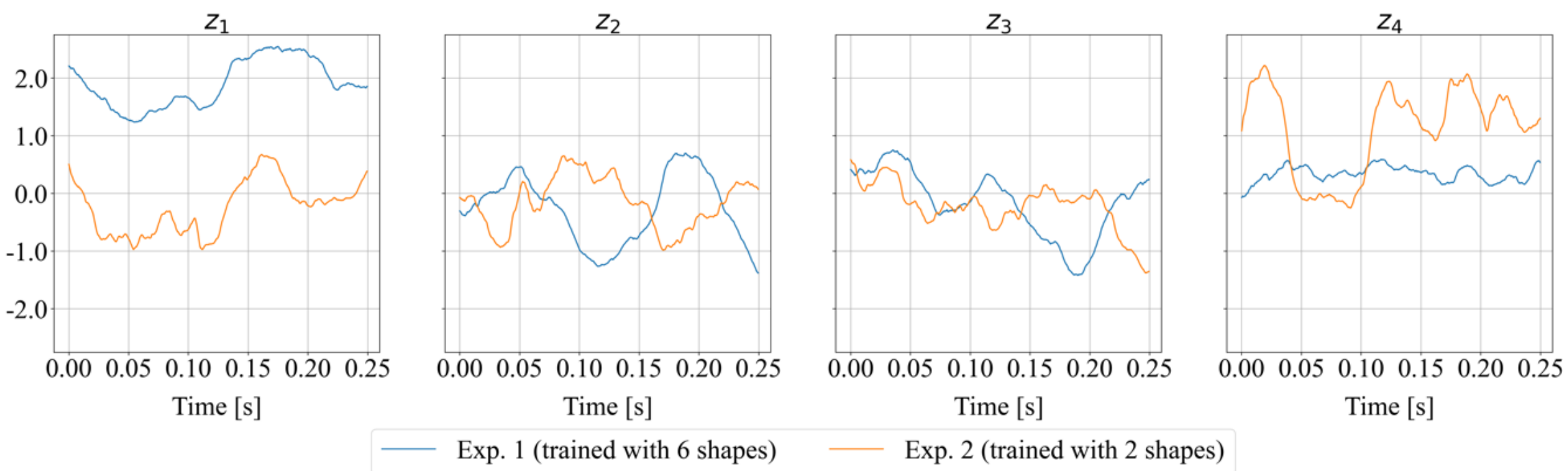

**Figure 7.** Representative temporal evolution of latent vector components $z_1$–$z_4$ for PC1 = 0.0. The trajectories compare Exp. 1 (blue, trained with 6 shapes) and Exp. 2 (orange, trained with 2 shapes), highlighting differences in amplitude and temporal patterns across latent dimensions.

Figure 8 presents a snapshot of the fluctuation component of the decomposed flow field, consisting of the three directional components of the flow velocity vector, corresponding to the first mode, i.e., $\boldsymbol{\mathcal{F}}_{dec}(\mu_1(t); \boldsymbol{\theta}_1)$ in Equation (10).

In Experiment 1, the flow field structure exhibits a more spatially continuous distribution compared to Experiment 2, although it cannot be regarded as fully physically natural. In contrast, the result of Experiment 2 shows a more pronounced spatial discontinuity. This is likely because Experiment 1 was trained using flow fields around a greater variety of geometries than Experiment 2, resulting in a more accurate representation of the global flow structures.

For the decomposed flow field corresponding to all modes, see Figure B.2 and B.3 in Appendix B.

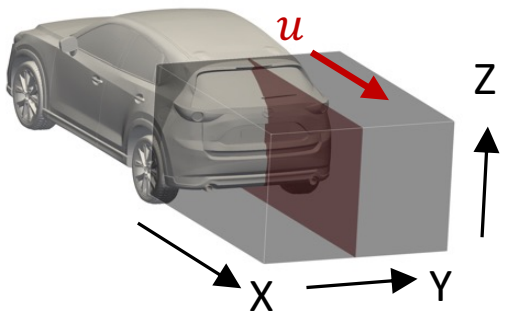

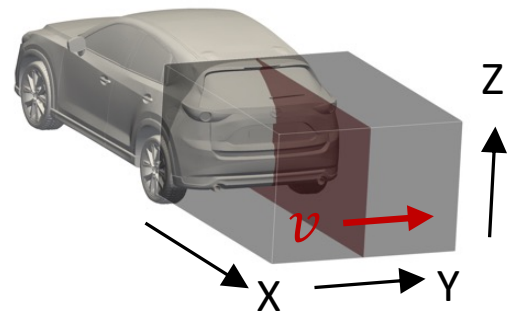

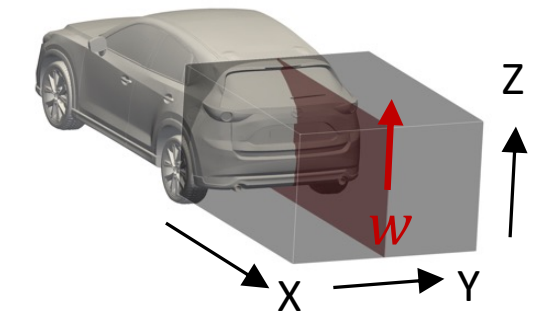

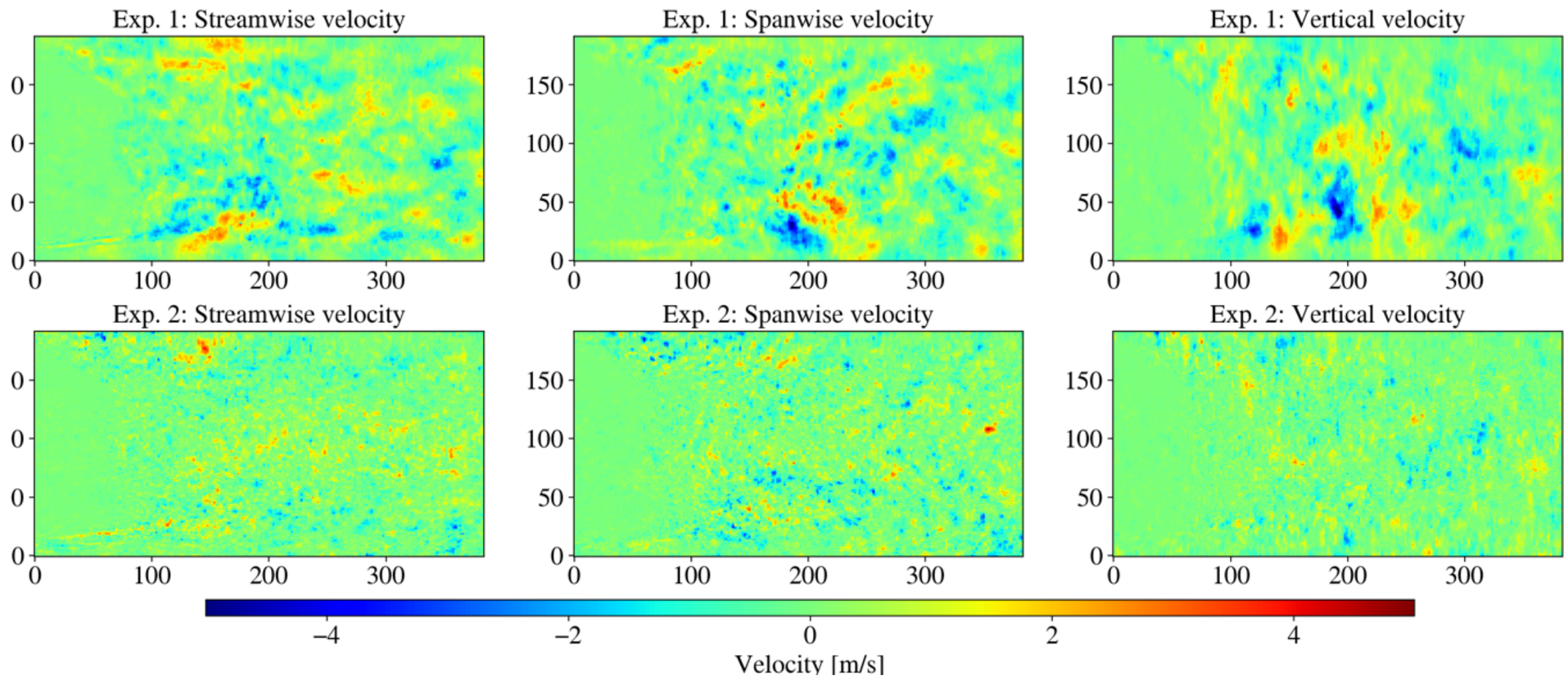

**Figure 8.** Snapshots of the fluctuating components of the decomposed flow field at PC1 = 0.0. The streamwise (u), spanwise (v), and vertical (w) velocity components are shown for Exp. 1 (top row) and Exp. 2 (bottom row). The schematics above indicate the coordinate system and velocity directions.

### *3.3. Evaluation of Model Reduction Accuracy*

This section evaluates the accuracy of the model reduction, i.e., how well the original flow field—which was not used for training—can be reconstructed from 32 latent variables using the proposed method.

First, the reconstructed flow fields, including the three directional components of the flow velocity vector, $\tilde{\boldsymbol{X}}(t)$ in Equation (10), are compared between Experiment 1 and Experiment 2. The vorticity in the y-direction ($\omega_y$ ) and velocity divergence, computed from this velocity field, are also analyzed. As an example, the flow field around shape PC1=0.0 is shown in Figure 9.

For the flow velocity field, both Experiment 1 and Experiment 2 successfully reproduce the overall flow structure observed in the reference high-precision simulation results. However, in Experiment 2, as observed in the decomposed flow field described earlier, the spatial continuity of the flow velocity is lost, resulting in a physically less realistic flow field. This suggests that Experiment 1 achieves higher reduction accuracy for the velocity field.

On the other hand, in the vorticity field, both Experiment 1 and Experiment 2 reproduce the clockwise vortex near the roof of the car body and the counterclockwise vortex near the bottom. However, it appears that the absolute value of the vorticity is more accurately reproduced in Experiment 2 than in Experiment 1. Although Experiment 2 shows higher accuracy in terms of vorticity magnitude, this does not necessarily indicate better generalization performance. Since Experiment 2 is trained on a smaller set of geometries, the model tends to specialize in representing specific flow structures, resulting in sharper vortical features. In contrast, Experiment 1, trained on a wider range of geometries, captures more generalized flow behavior, which leads to smoother but slightly underestimated vorticity fields. This suggests a trade-off between generalization capability and the accuracy of local flow structures.

The spatial distribution of velocity divergence is also shown on the right side of Figure 9. In the reference simulation results, the divergence remains close to zero throughout the domain, whereas in Experiment 1 it deviates from zero with both positive and negative values. In Experiment 2, the magnitude of this deviation is further increased compared to Experiment 1. These results indicate that, while the reference simulation satisfies mass conservation, the flow fields reconstructed in Experiments 1 and 2 violate mass conservation. However, in a relative sense, Experiment 1 exhibits more physically consistent behavior than Experiment 2. For the complete flow fields around other shapes, see Figures C.1–C.4 in Appendix C.

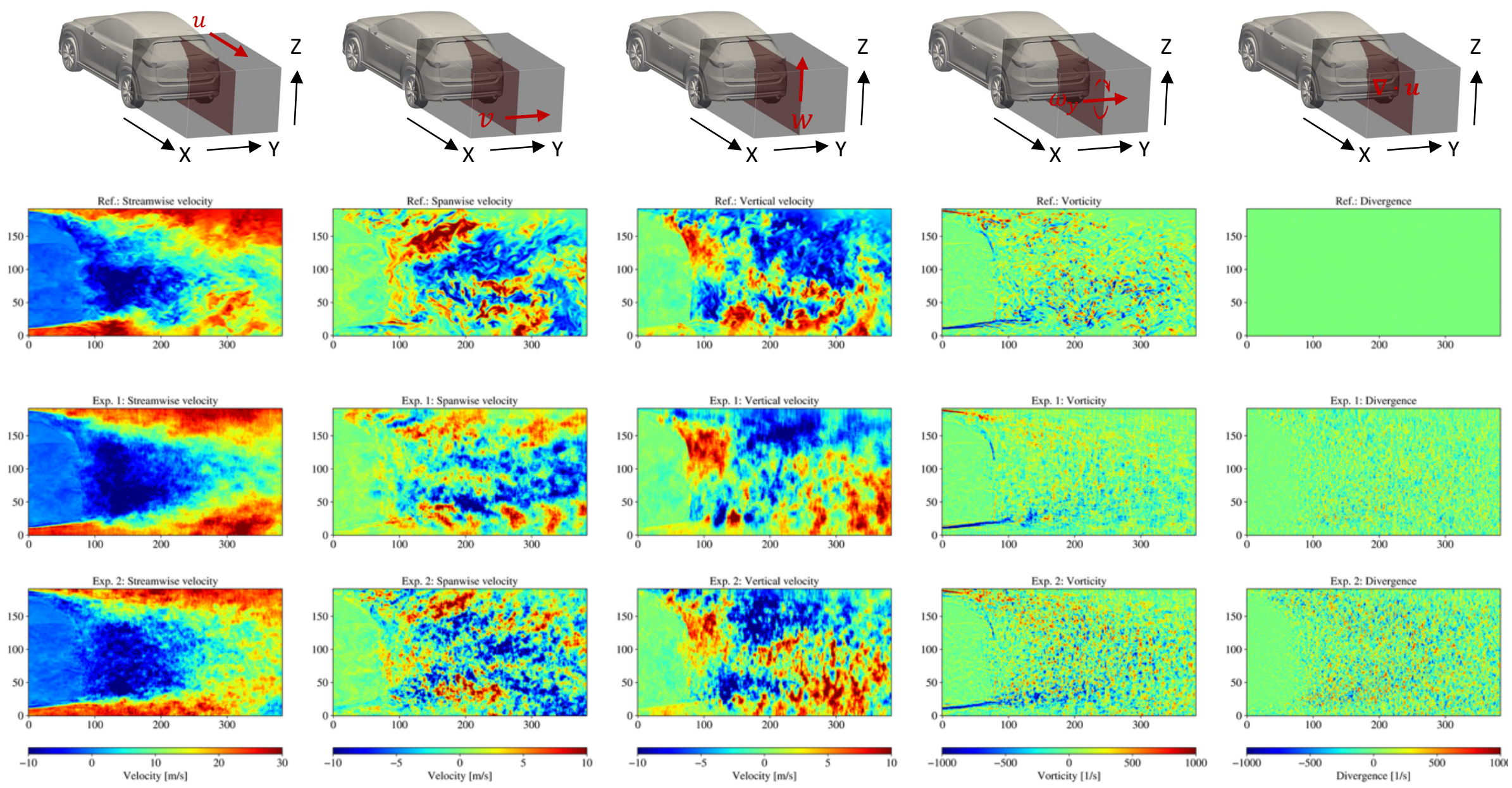

**Figure 9.** Snapshots of reconstructed flow fields from 32 latent variables at PC1 = 0.0. The streamwise (u), spanwise (v), vertical (w) velocity components, vorticity, and velocity divergence are shown for the reference field (top row), Exp. 1 (middle row), and Exp. 2 (bottom row). The schematics above indicate the coordinate system and velocity directions.

To confirm the three-dimensional structure of the vortex, the instantaneous flow field of the Q-criterion (the second invariant of the velocity gradient tensor) isosurface for PC1=0.0 is shown in Figure 10. The Q-criterion is formulated as follows:

$$Q = \frac{1}{2}(\|\Omega\|^2 - \|S\|^2), \tag{10}$$

where $\Omega$ and $S$ represent the rate of rotation tensor and the rate of strain tensor, respectively.

The results of Experiment 1 indicate that the three-dimensional vortex structures observed in the high-precision simulation results are qualitatively reproduced. In particular, the large-scale coherent wake structures and their spatial distributions show good agreement with the reference field, suggesting that the dominant flow characteristics are successfully captured. In contrast, the results of Experiment 2 deviate from the high-precision simulation results, as the vortices appear finer in scale and are more uniformly distributed throughout the computational domain. This suggests that small-scale fluctuations are overrepresented, while the large-scale coherent structures are less accurately reproduced. This difference is likely because Experiment 1 was trained on flow fields around a larger number of geometries, allowing it to learn the global characteristics of the turbulent wake more effectively.
For the Q-criterion isosurfaces around other shapes, see Figures D.1–D.4 in Appendix D.

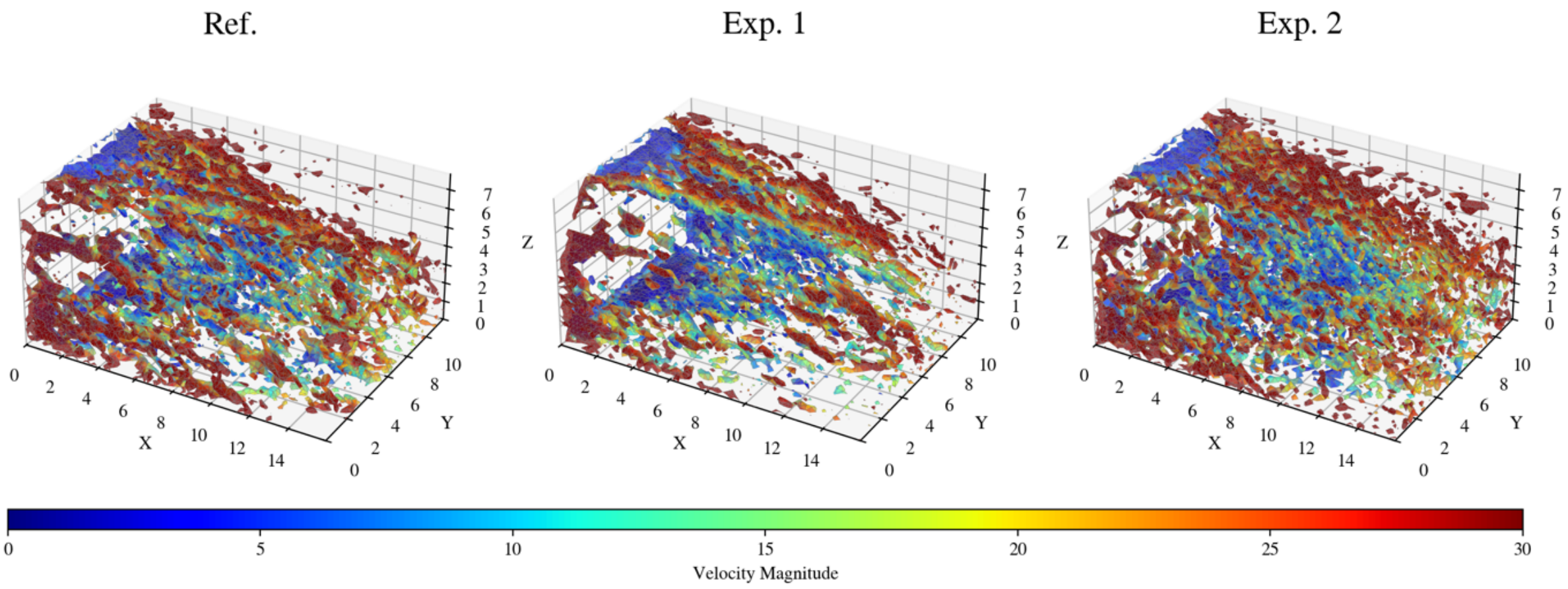

**Figure 10.** Instantaneous Q-criterion isosurfaces colored by velocity magnitude at PC1 = 0.0 for the reference field (left), Exp. 1 (center), and Exp. 2 (right).

Figure 11 illustrates the energy spectrum at the probe point located at the rear end of the vehicle body. In Experiment 1, the amplitude of the high-frequency components more closely matches the high-precision simulation results than in Experiment 2. This indicates that Experiment 1 achieves higher reproduction accuracy for eddies with small time scales compared to Experiment 2. Furthermore, Experiment 1 reproduces the overall spectral distribution more consistently across a wide frequency range. In contrast, Experiment 2 exhibits a more rapid attenuation in the high-frequency region, indicating a loss of small-scale turbulent fluctuations. These results suggest that Experiment 1 better preserves the temporal characteristics of the turbulent wake and captures a broader range of turbulent scales.

For the spectra corresponding to other shapes, see Figure E.1 in Appendix E.

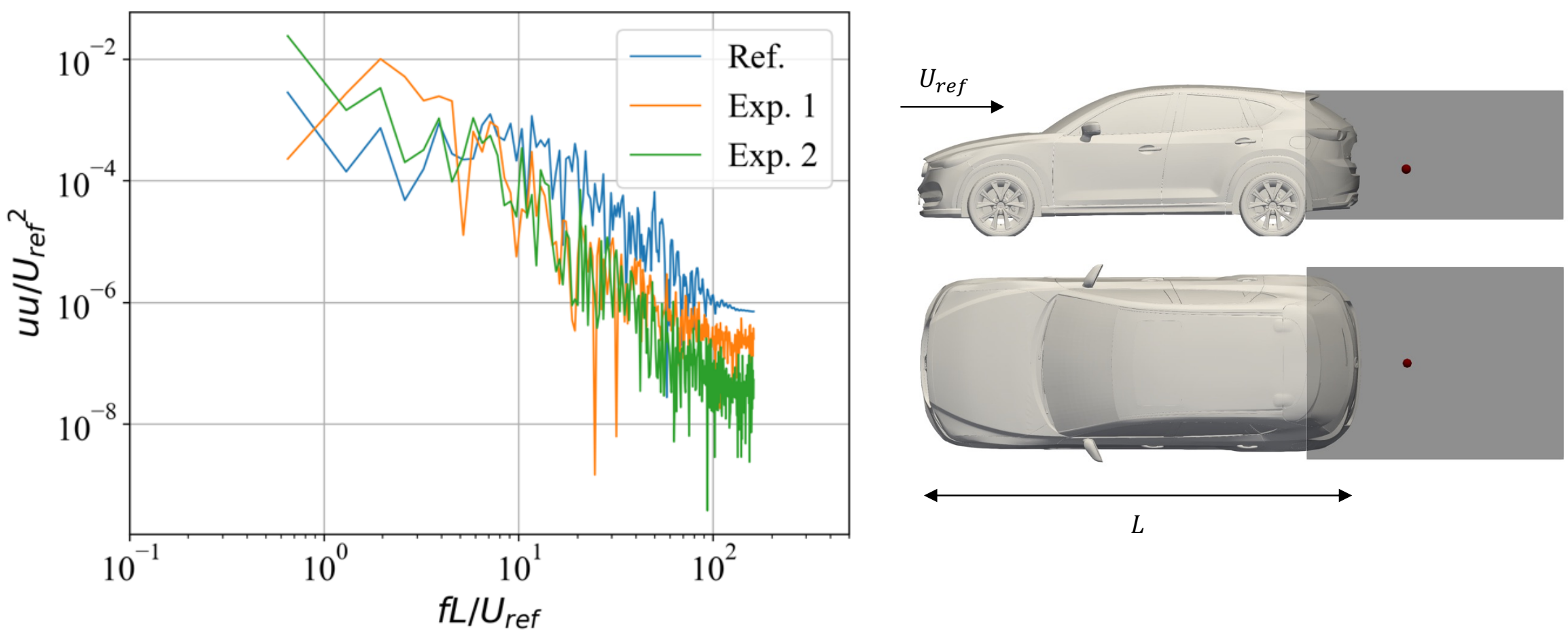

**Figure 11.** Comparison of energy spectra at a specific probe point in the vehicle wake. (Left) The plot shows the non-dimensional energy spectrum ($uu/U_{ref}^2$) against the non-dimensional frequency ($fL/U_{ref}$) for the reference case and two experiments. (Right) The schematic illustrates the location of the measurement probe (red dot) in the wake region behind the vehicle model, defining the reference velocity $U_{ref}$ and vehicle length $L$

Figure 12 illustrates the probability density function (PDF) of vorticity for PC1 = 0.0. For $\omega_y$, both Experiment 1 and Experiment 2 show reduced reproduction accuracy in regions with large absolute values, corresponding to intense small-scale vortical structures. This suggests that strong and localized vortices are more difficult to reconstruct than moderate vorticity regions.

In contrast, for $\omega_z$, Experiment 2 shows good agreement with the reference high-precision simulation results even in regions with large absolute values. This is likely because the vehicle geometries vary mainly along the longitudinal direction, while little variation exists in the left–right direction. As a result, the $\omega_z$ distribution is less sensitive to shape differences than $\omega_y$, leading to similar statistical characteristics across different vehicle configurations.

These results indicate that the prediction accuracy depends not only on the magnitude of vorticity but also on the degree of geometric variability associated with each flow component.
For the PDFs corresponding to other shapes, see Figure F.1 in Appendix F.

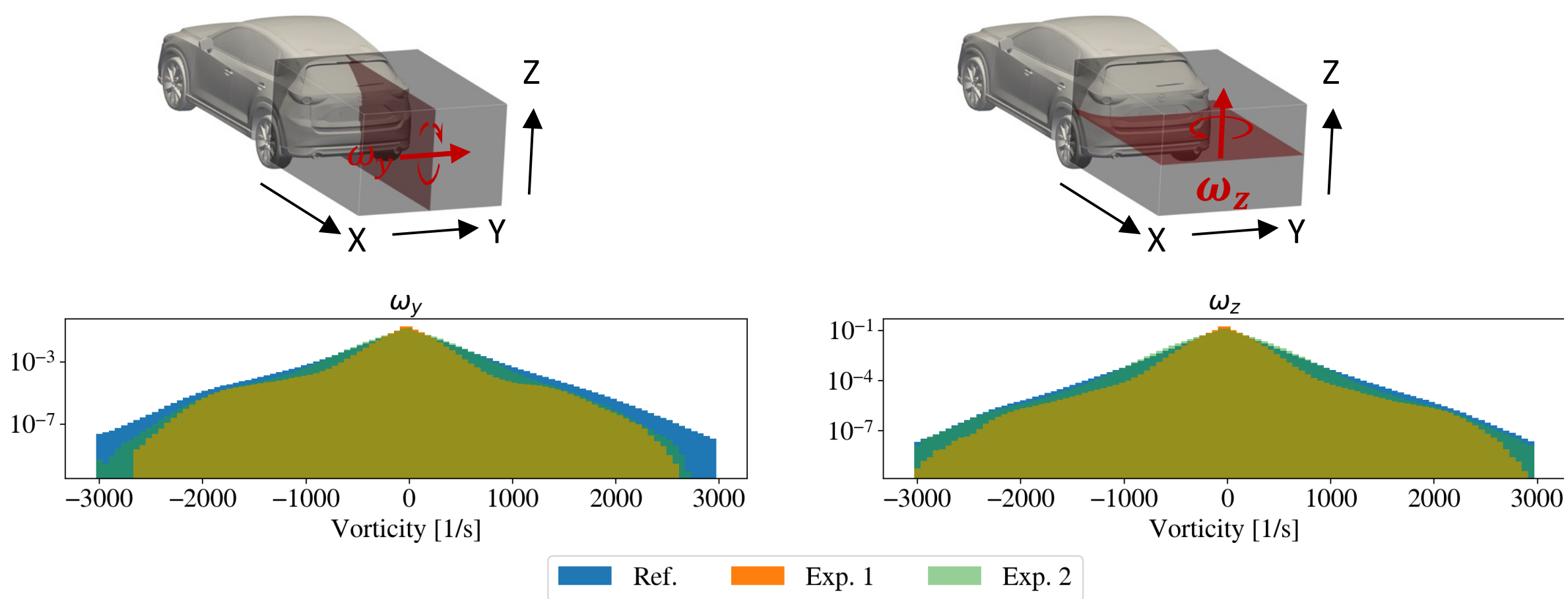

**Figure 12.** Probability density functions (PDFs) of vorticity in the vehicle wake. The top schematics define the analysis planes for the y-component ($\omega_y$, left) and z-component ($\omega_z$, right) of vorticity within the wake region behind the vehicle body. The bottom plots show the corresponding PDFs, comparing the reference (Ref.) and two experimental setups (Exp. 1 and Exp. 2).

Figure 13 illustrates the trajectory of a physical system in phase space, spanned by three variables: turbulent kinetic energy (TKE), TKE production, and TKE dissipation for PC1=0.0. These quantities are formulated as follows:

Turbulent Kinetic Energy (TKE):

$$\int_V k dV = \int_V \frac{1}{2}\left(u'^2 + v'^2 + w'^2\right) dV, \quad (11)$$

Production of TKE:

$$\int_V P dV = \int_V -u' v' \frac{dU}{dy} dV, \quad (12)$$

Dissipation of TKE:

$$\int_V D dV = \int_V \nu\left(\|\nabla u'\|^2 + \|\nabla v'\|^2 + \|\nabla w'\|^2\right) dV, \quad (13)$$

where $k$ represents the TKE, $P$ is the TKE production rate, $D$ is the TKE dissipation rate, $V$ is the target training region, and $U$ is the mean streamwise velocity.

In both Experiment 1 and Experiment 2, the trajectories in phase space deviate from those of the high-precision simulation results, indicating that the balance among turbulent kinetic energy production, transport, and dissipation is not fully reproduced. In particular, the trajectories form distinct clusters separated from the reference field in the $P$-$k$, $P$-$D$, and $k$-$D$ planes, suggesting differences in the energetic state of the reconstructed turbulent flow.

Experiment 1 shows closer agreement with the reference results than Experiment 2, whereas Experiment 2 exhibits larger deviations, particularly in the TKE production and dissipation terms. These results suggest that although the proposed approach can reproduce major flow structures, accurately preserving the energy transfer process across turbulent scales remains challenging.

For the time variation of TKE corresponding to other shapes, see Figure G.1 in Appendix G.

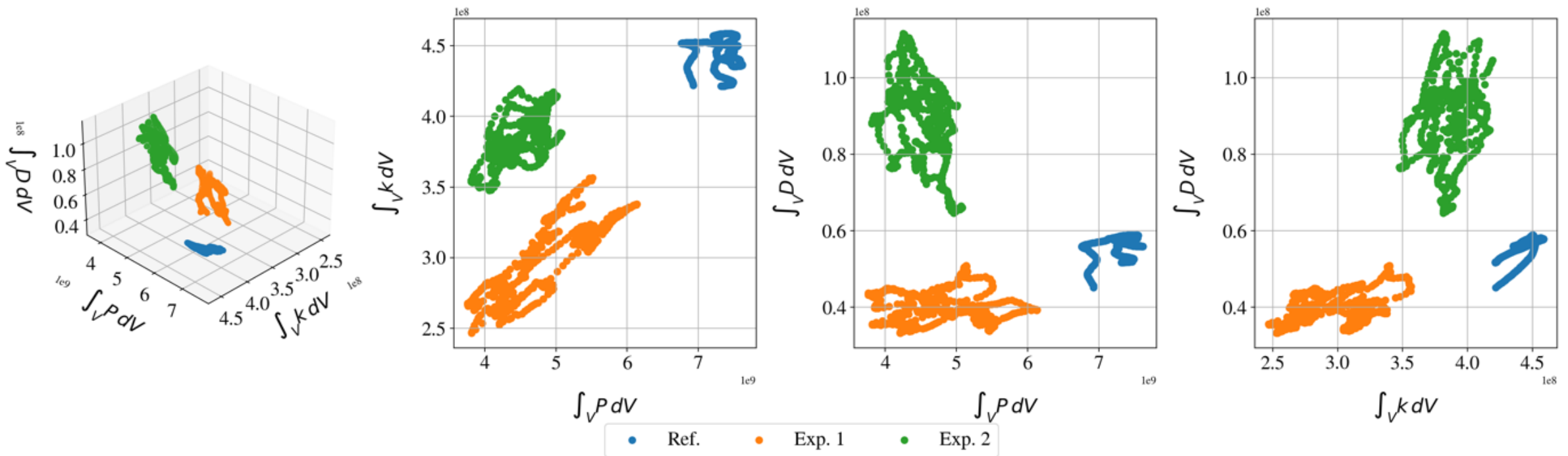

**Figure 13.** Comparison of phase space trajectories for different experimental setups. The phase space is defined by the turbulent kinetic energy (TKE, $\int_V k\,dV$), TKE production ($\int_V P\,dV$), and TKE dissipation ($\int_V D\,dV$). From left to right, the panels display a 3D perspective view followed by 2D projections onto the $P$-$k$, $P$-$D$, and $k$-$D$ planes. Distinct clusters for Ref. (blue), Exp. 1 (orange), and Exp. 2 (green) illustrate the shifts in the system's energetic state under each condition.

Figure 14 illustrates the temporal correlation coefficient for PC1=0.0. The temporal correlation coefficient is formulated as follows:

$$R(\Delta t) = \frac{\langle u'(t)u'(t + \Delta t)\rangle}{\langle u'(t)^2\rangle}, \tag{14}$$

where $R$ represents the temporal correlation coefficient, $\langle\cdots\rangle$ denotes the spatial average, $u'$ is the fluctuation component of the streamwise velocity, $t$ denotes the initial time, and $\Delta t$ is the time interval from the initial state.

Regarding the time variation of the temporal correlation coefficient, the results of Experiment 1 are closer to the high-precision simulation results than those of Experiment 2. In particular, Experiment 1 reproduces the initial decay behavior of the temporal correlation more accurately, indicating better agreement in the characteristic time scale of the turbulent fluctuations.

However, both Experiment 1 and Experiment 2 retain larger correlation values than the reference results at later times, suggesting that temporal decorrelation occurs more slowly in the reconstructed flow fields. This indicates that the generated flow fields contain more persistent structures and do not fully reproduce the temporal evolution of turbulent fluctuations.

These results suggest that Experiment 1 more accurately captures the temporal characteristics of the turbulent wake, although reproducing the complete temporal dynamics of turbulence remains challenging.

For the correlation coefficients corresponding to other shapes, see Figure H.1 in Appendix H.

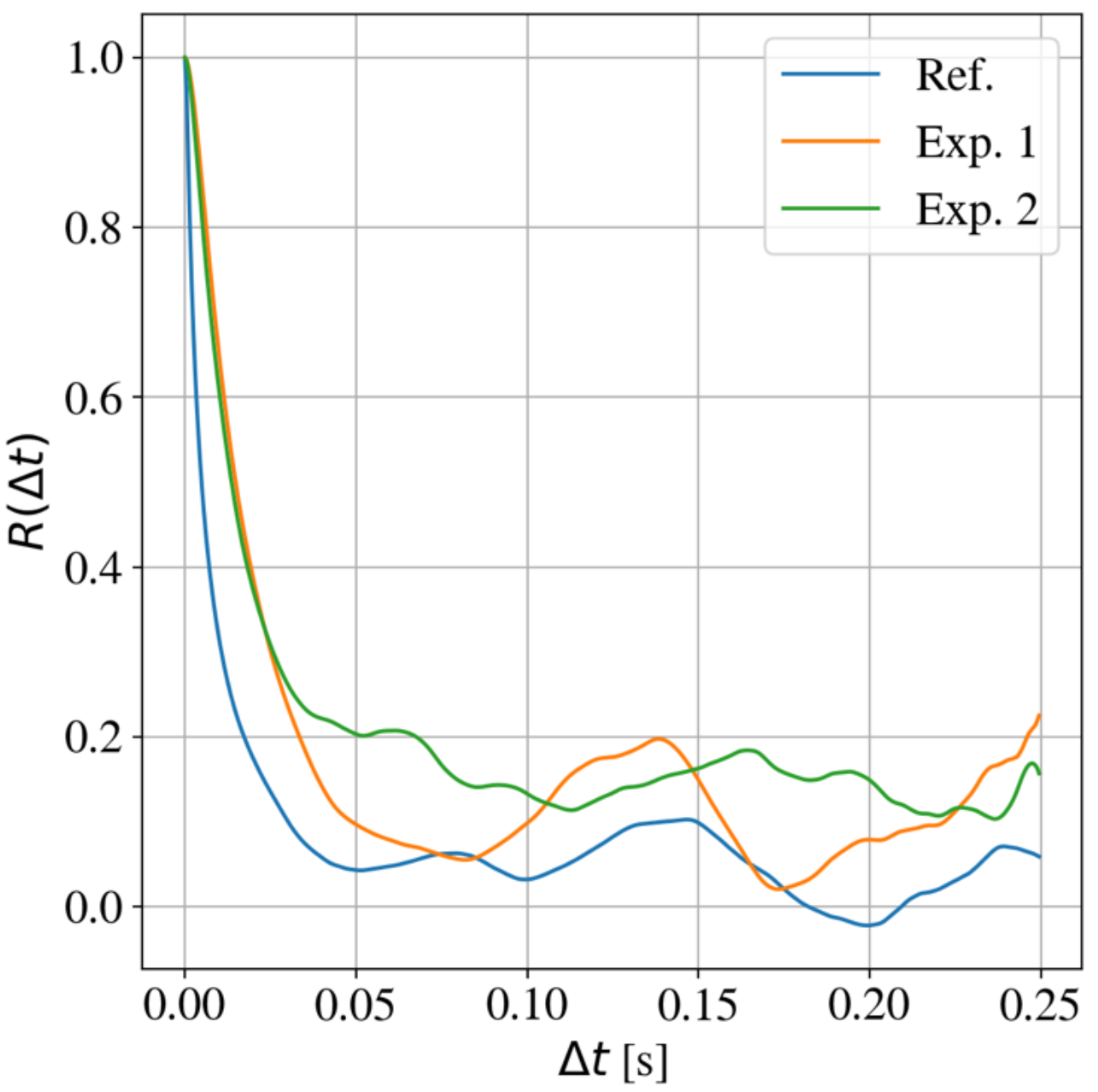

**Figure 14.** Temporal autocorrelation coefficient $R(\Delta t)$ of velocity fluctuations for PC1=0.0. The plot compares the decay of temporal correlation over a time interval $\Delta t$ [s] for the Ref. (blue), Exp. 1 (orange), and Exp. 2 (green).

### *3.4. Computational Speed-Up via Model Reduction*

The computational speed-up achieved by the proposed model reduction method during simulation is presented. As shown in Table 3, the proposed method requires up to 133 hours of computation for training. However, once the training is completed, the reduced-order simulation for arbitrary geometries only requires the time evolution of 32 variables to be computed. The computational cost for this is $1.43\times10^{-3}$ seconds per time step (corresponding to an integration time of $2.5\times10^{-5}$ seconds) using a single CPU.

In contrast, the high-precision simulation requires 385 CPUs, with a computational time of 5.69 seconds per time step. This corresponds to a performance improvement of approximately 3,979 times in terms of computational time.

## 4. Summary and conclusions

In this study, we applied a new variational autoencoder (VAE) mechanism to our distributed parallel training-based model reduction framework for high-precision flow field data, developed in our previous work. We focused on the region near the rear end of the vehicle body to evaluate its robustness in handling high-Reynolds-number flows around unknown vehicle shapes.

Specifically, we designed two experiments, where the flow fields around either five or two vehicle shapes were used for training. We then evaluated whether vortex generation at various spatial and temporal scales could be reconstructed from 32 latent variables for flow fields around shapes not included in training. In the largest experiment, training was conducted for 26 hours using 8,118 nodes on Fugaku.

As a result, both experiments demonstrated high accuracy in reproducing large-scale vortices in space and time. However, a clear trade-off between generalization capability and the accuracy of local flow structures was observed. Experiment 1, trained on a larger number of geometries, successfully captured more generalized flow behavior. It exhibited superior physical consistency, demonstrated by more spatially continuous velocity fields , better adherence to mass conservation (lower velocity divergence) , and closer agreement with reference high-precision simulations in the turbulent kinetic energy (TKE) phase space. Furthermore, Experiment 1 more accurately preserved the initial temporal correlation decay and the high-frequency components of the energy spectrum.

In contrast, Experiment 2, trained on fewer geometries, tended to specialize in specific flow patterns. While this allowed for a sharper representation of strong, localized vortical structures, particularly in the z-direction ($\omega_z$) , it also resulted in pronounced spatial discontinuities and an overrepresentation of fine-scale, physically inconsistent fluctuations.

Notably, the proposed model reduction method achieved significant computational efficiency. Once the network is trained, the reduced-order simulation for arbitrary geometries requires computing the time evolution of only 32 variables, taking just $1.43\times10^{-3}$ seconds per time step on a single CPU. This represents a computational speed-up of approximately 3,979 times compared to the high-precision simulation requiring 385 CPUs.

For future work, we aim to enhance the network architecture to further improve the accuracy of small-scale vortex reconstruction and optimize hyperparameters, including determining the appropriate number of decomposition modes based on vortex scale.

**Acknowledgments**
This research is an extension of the method originally proposed by Mr. Takaaki Murata, Dr. Kai Fukami, and Professor Koji Fukagata of Keio University. We sincerely appreciate their valuable contributions, which laid the foundation for this study, as well as their insightful advice throughout the research.

We also extend our gratitude to Professor Takuji Nakashima from Hiroshima University for providing high-quality vehicle body models and for engaging in fruitful discussions.

The computational resources of "Fugaku" were provided through the HPCI System Research Project (Project ID: hp240205 and ra240008).

**Appendices**

*Appendix A*

The detailed network architectures for the encoder and decoder are presented in Tables A.1 and A.2

**Table A.1.** Layer-wise details of the encoder network architecture.

| Layer type | Output shape (X, Y, Z, channel) |
|---|---|
| Input layer | (384, 288, 192, 3) |

| | | |
|---|---|---|
| 1st | 3D convolutional layer | (384, 288, 192, 4) |
| | Layer normalization | (384, 288, 192, 4) |
| | 3D max pooling layer | (192, 144, 96, 4) |
| 2nd | 3D convolutional layer | (192, 144, 96, 2) |
| | Layer normalization | |
| | 3D max pooling layer | (96, 72, 48, 2) |
| 3rd | 3D convolutional layer | (96, 72, 48, 2) |
| | Layer normalization | (96, 72, 48, 2) |
| | 3D max pooling layer | (48, 36, 24, 2) |
| 4th | 3D convolutional layer | (48, 36, 24, 1) |
| | Layer normalization | (48, 36, 24, 1) |
| | 3D max pooling layer | (24, 18, 24, 1) |
| 5th | 3D convolutional layer | (24, 18, 24, 3) |
| | Layer normalization | (24, 18, 24, 3) |
| | 3D max pooling layer | (12, 9, 24, 3) |
| 6th | 3D convolutional layer | (12, 9, 24, 3) |
| | Layer normalization | (12, 9, 24, 3) |
| | 3D max pooling layer | (6, 4, 24, 3) |
| 7th | Fully connected layer | (32, 1, 1, 1) |
| | Layer normalization (latent vector output) | (32, 1, 1, 1) |

**Table A.2.** Layer-wise details of the decoder network architecture.

| Layer type | | Output shape (X, Y, Z, channel) |
|---|---|---|
| Initial input (single element of the latent vector) | | (1, 1, 1, 1) |
| 1st | Fully connected layer | (6, 4, 24, 3) |
| | Layer normalization | (6, 4, 24, 3) |
| 2nd | 3D interpolation layer | (12, 9, 24, 3) |
| | 3D convolutional layer | (12, 9, 24, 3) |
| | Layer normalization | (12, 9, 24, 3) |
| 3rd | 3D interpolation layer | (24, 18, 24, 3) |
| | 3D convolutional layer | (24, 18, 24, 1) |
| | Layer normalization | (24, 18, 24, 1) |
| 4th | 3D interpolation layer | (48, 36, 24, 1) |
| | 3D convolutional layer | (48, 36, 24, 2) |
| | Layer normalization | (48, 36, 24, 2) |
| 5th | 3D interpolation layer | (96, 72, 48, 2) |
| | 3D convolutional layer | (96, 72, 48, 2) |
| | Layer normalization | (96, 72, 48, 2) |
| 6th | 3D interpolation layer | (192, 144, 96, 2) |
| | 3D convolutional layer | (192, 144, 96, 2) |
| | Layer normalization | (192, 144, 96, 2) |
| 7th | 3D interpolation layer | (384, 288, 192, 2) |
| | 3D convolutional layer | (384, 288, 192, 3) |
| | Layer normalization (decomposed flow field) | (384, 288, 192, 3) |

The hyperparameters used in this study are presented in Table A.3.

**Table A.3.** Hyperparameter settings.

| | |
|---|---|
| Filter size (convolutional layers) | $3 \times 3 \times 3$ |
| Pooling size (max pooling layers) | $2 \times 2 \times 2$ |
| Weight initialization method | Xavier uniform |
| Optimization algorithm | Adam |
| Initial learning rate | 0.001 |
| Batch size per iteration | 492 |
| Activation function | tanh |

*Appendix B*

The time-variation of elements of latent vector are shown in Figure B.1.

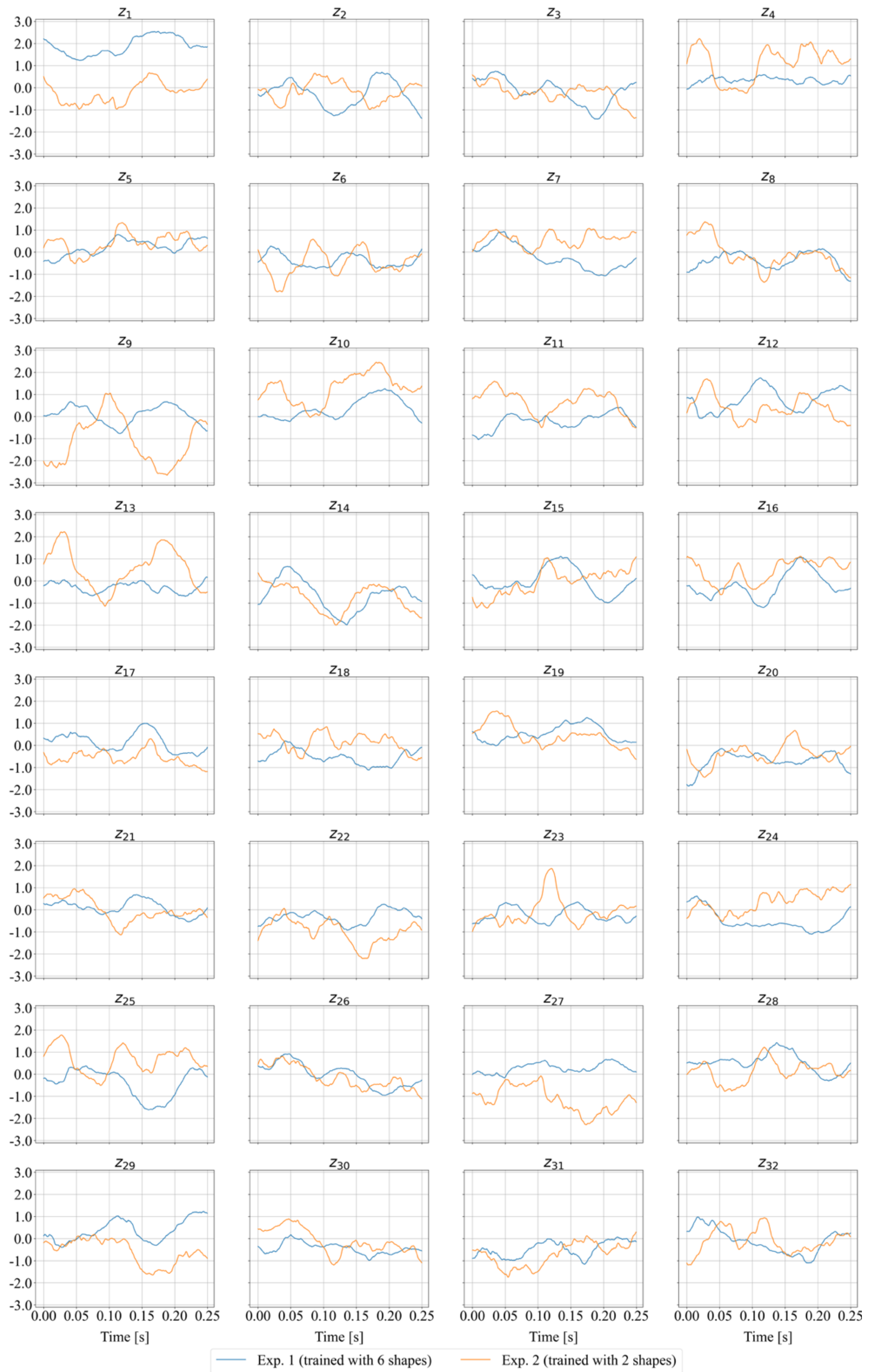

**Figure B.1.** Time variation of all latent vector elements for PC1=0.0.

Here, snapshots of the fluctuating components of the decomposed streamwise flow field around PC1=0.0 for each mode are presented. Figures B.2 and B.3 correspond to Experiment 1 and Experiment 2, respectively.

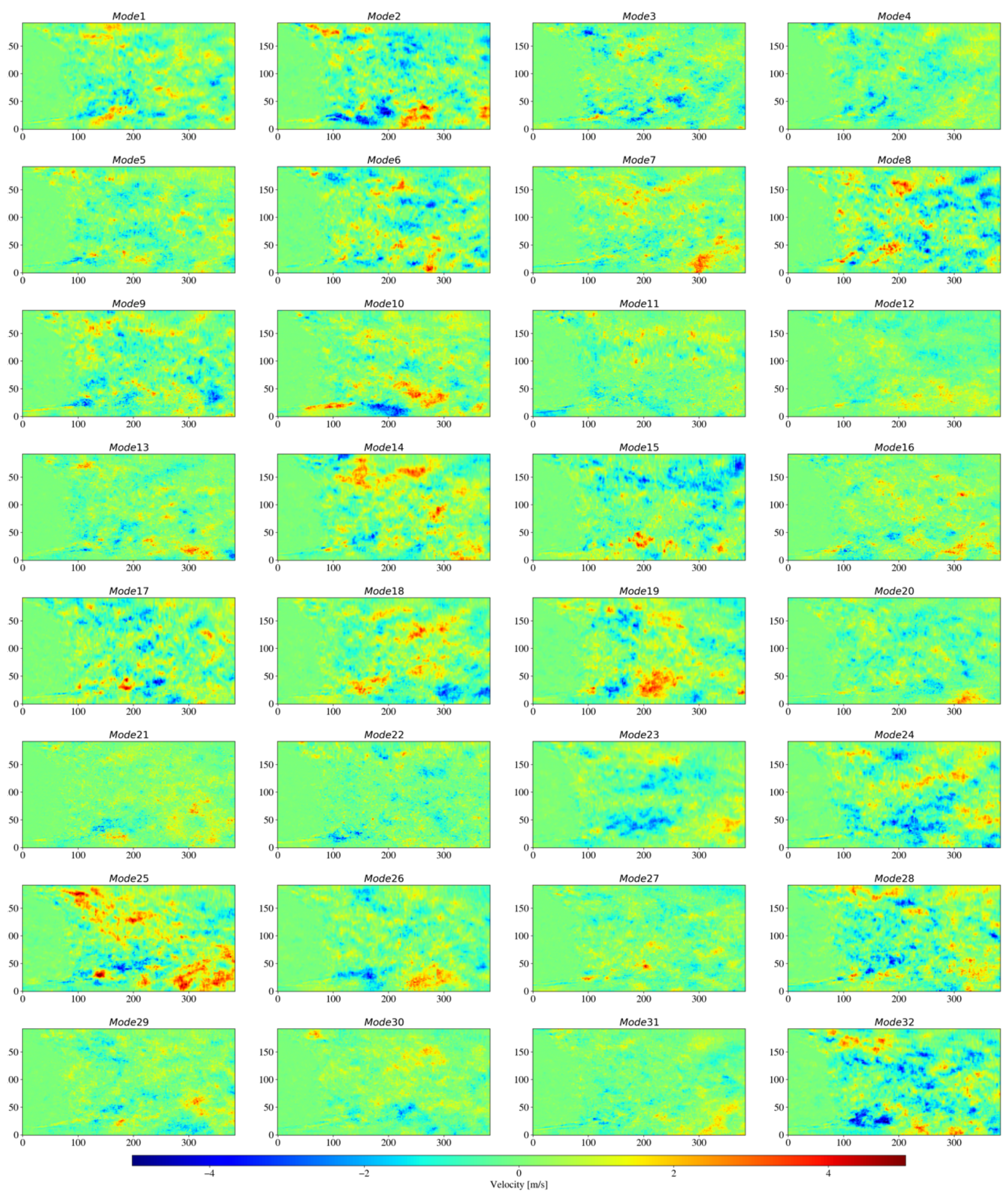

**Figure B.2.** Snapshot of the fluctuating components of the streamwise flow velocity $u$ for each mode (Exp. 1).

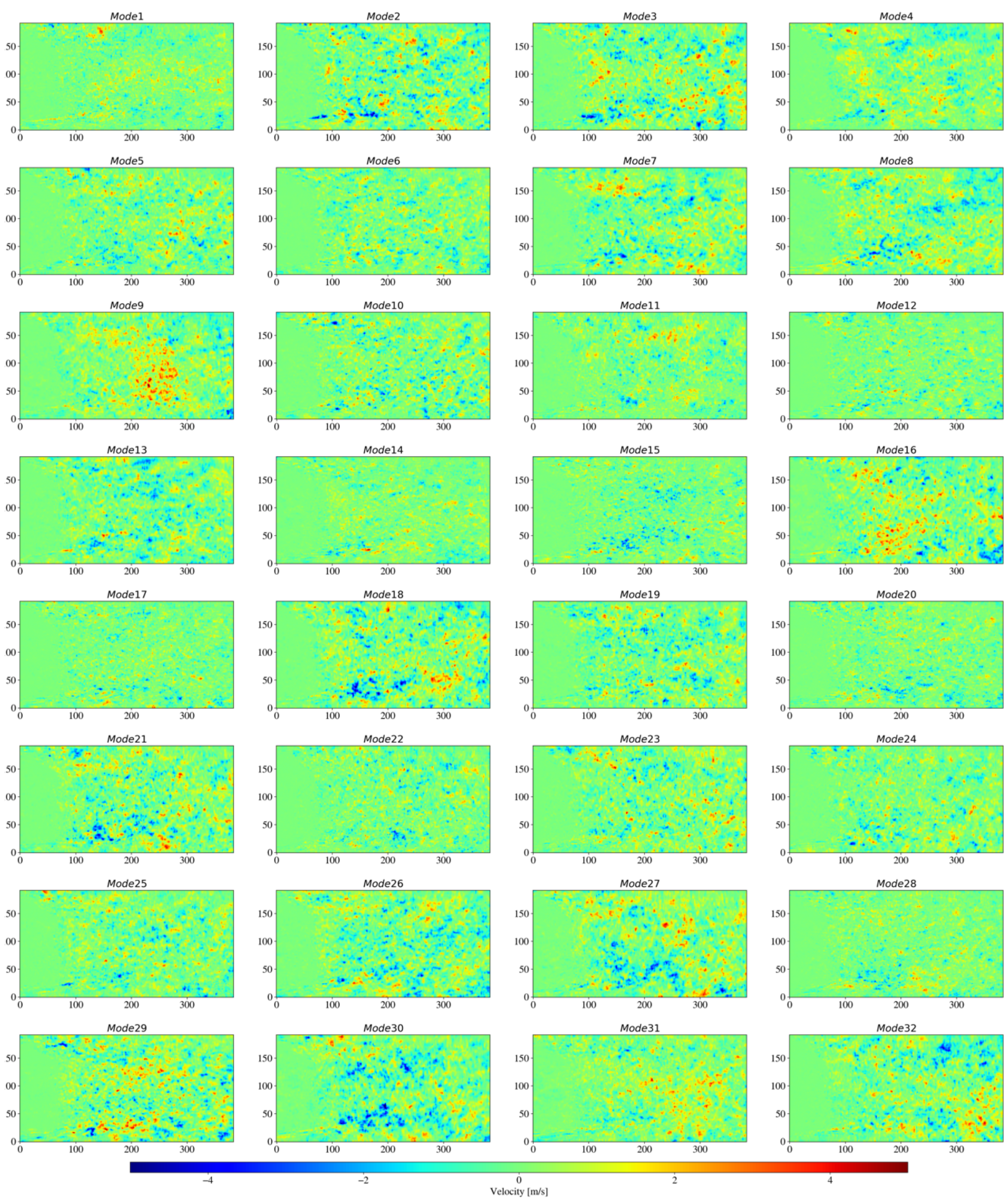

**Figure B.3.** Snapshot of the fluctuating components of the streamwise flow velocity $u$ for each mode (Exp. 2).

*Appendix C*

Figure C.1–C.4 presents a snapshot of the reconstructed flow fields from 32 latent variables for all vehicle geometries.

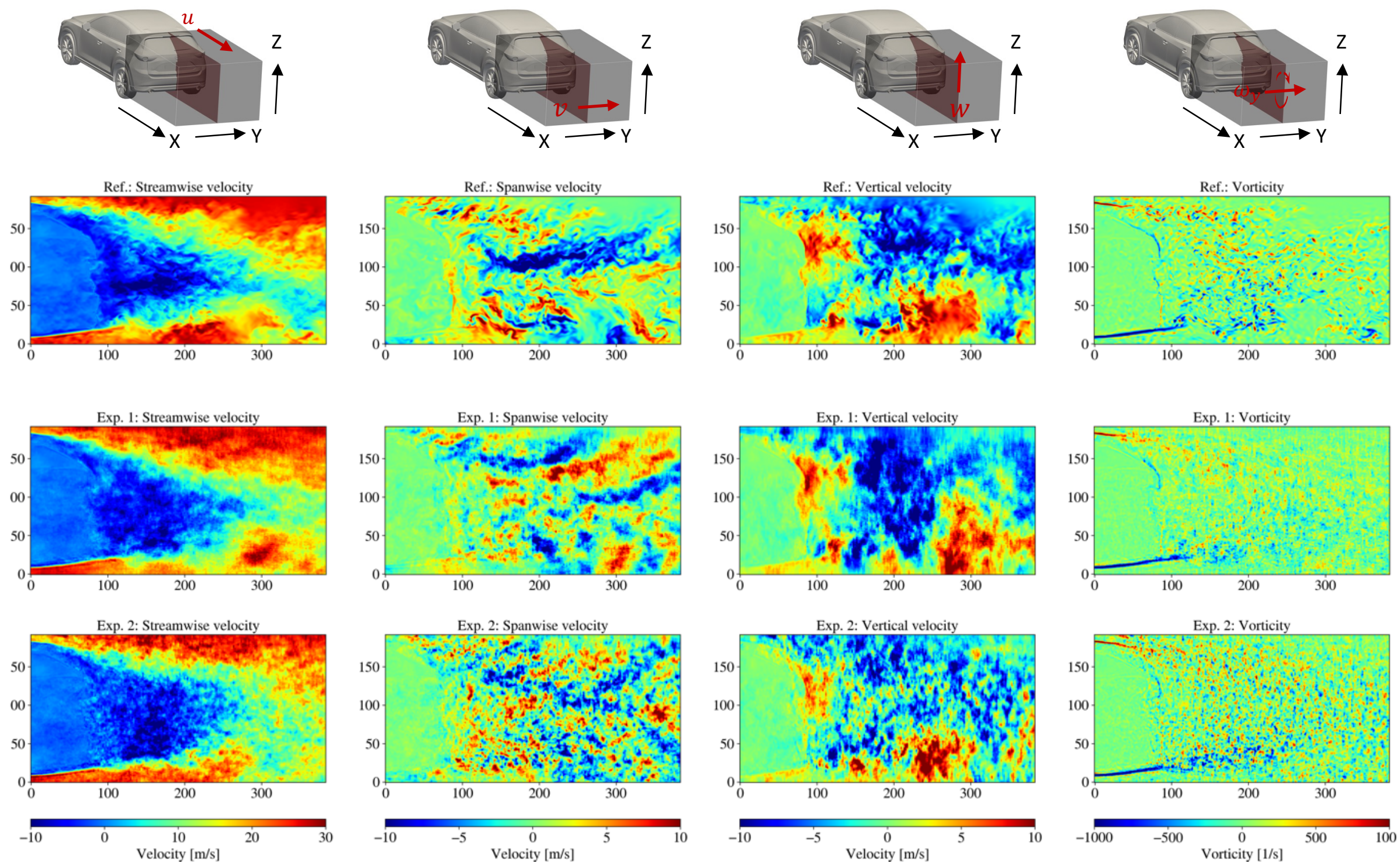

**Figure C.1** Snapshot of the reconstructed flow fields from 32 latent variables for PC1=-1.2.

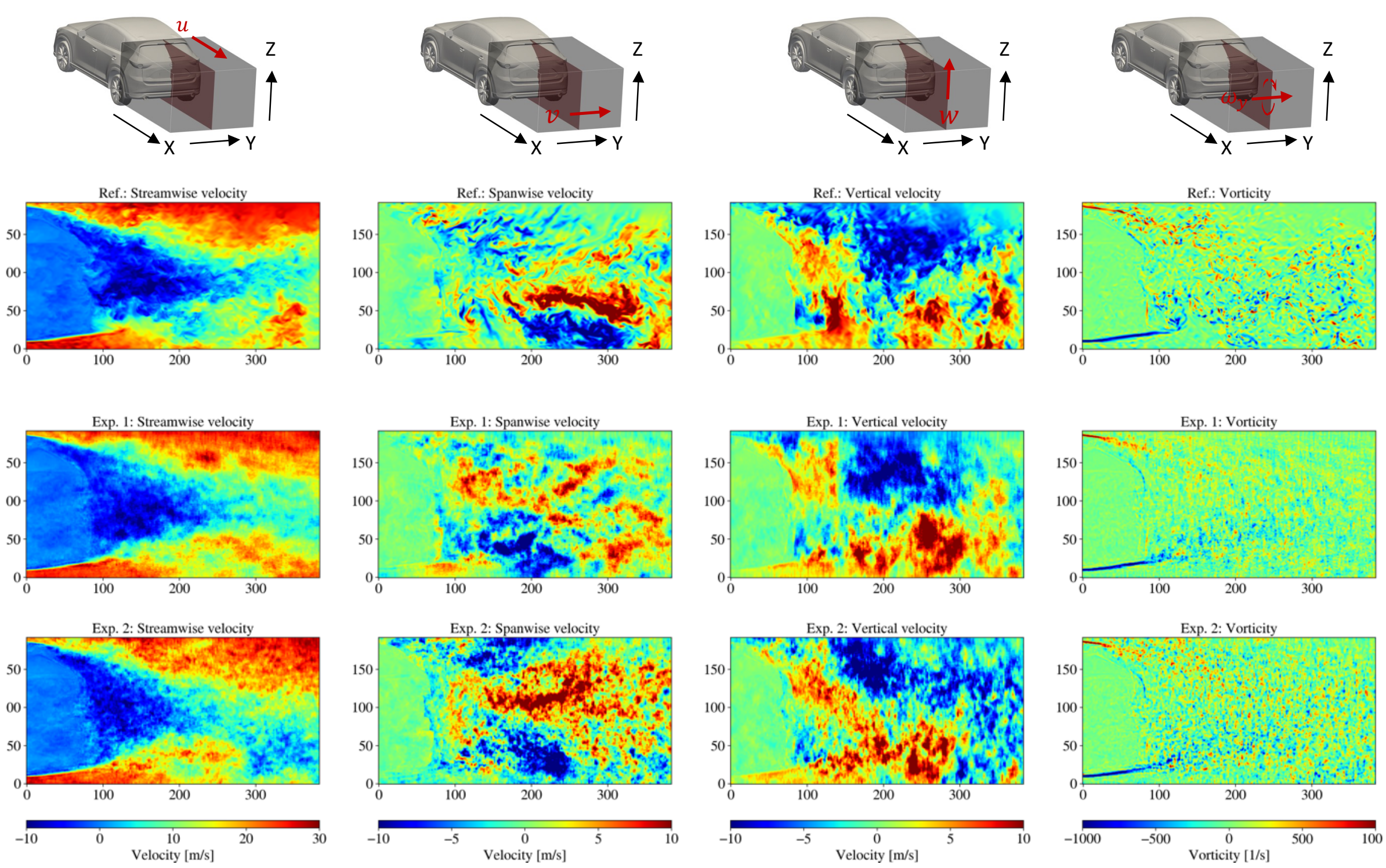

**Figure C.2** Snapshot of the reconstructed flow fields from 32 latent variables for PC1=-0.6.

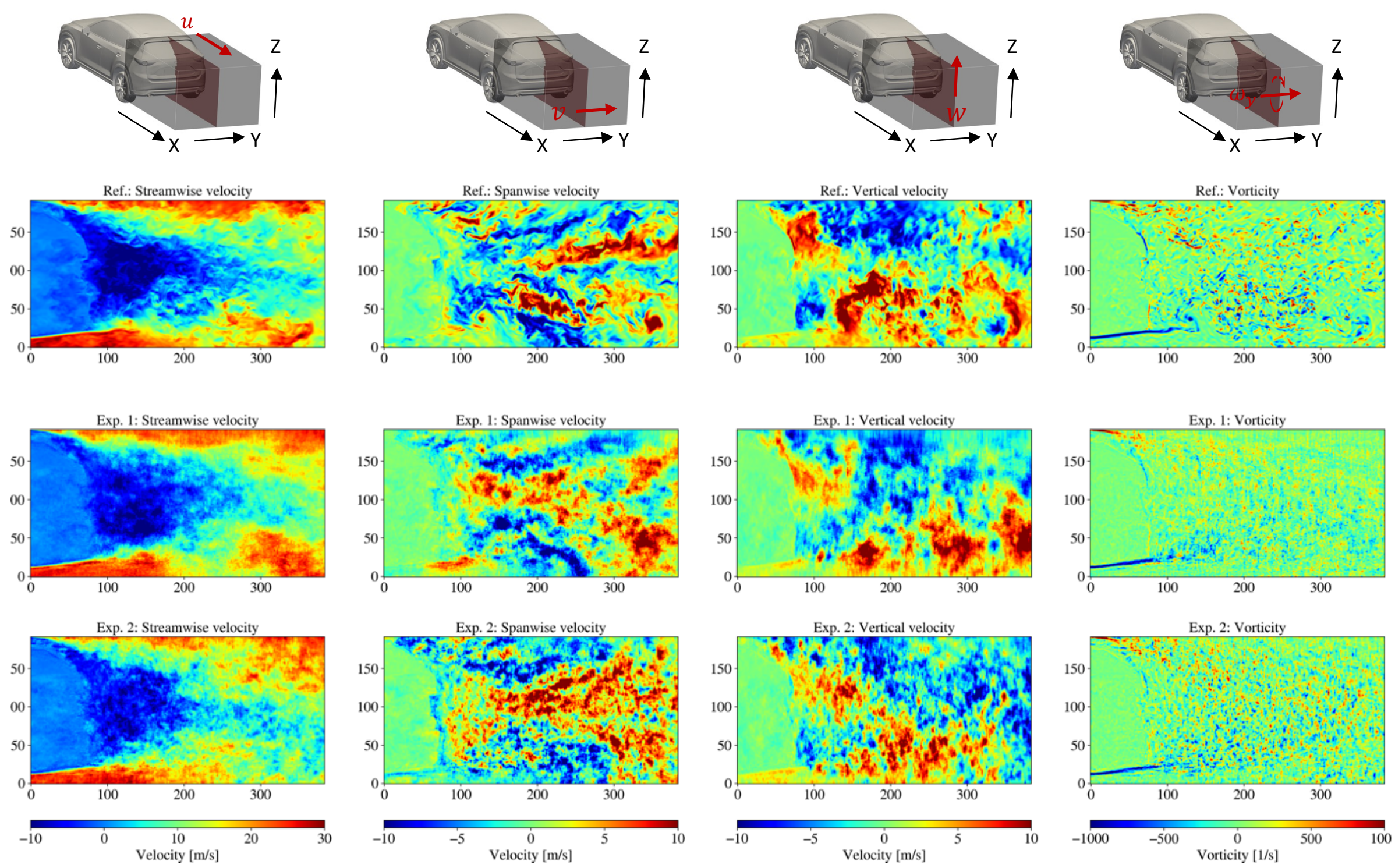

**Figure C.3** Snapshot of the reconstructed flow fields from 32 latent variables for PC1=0.6.

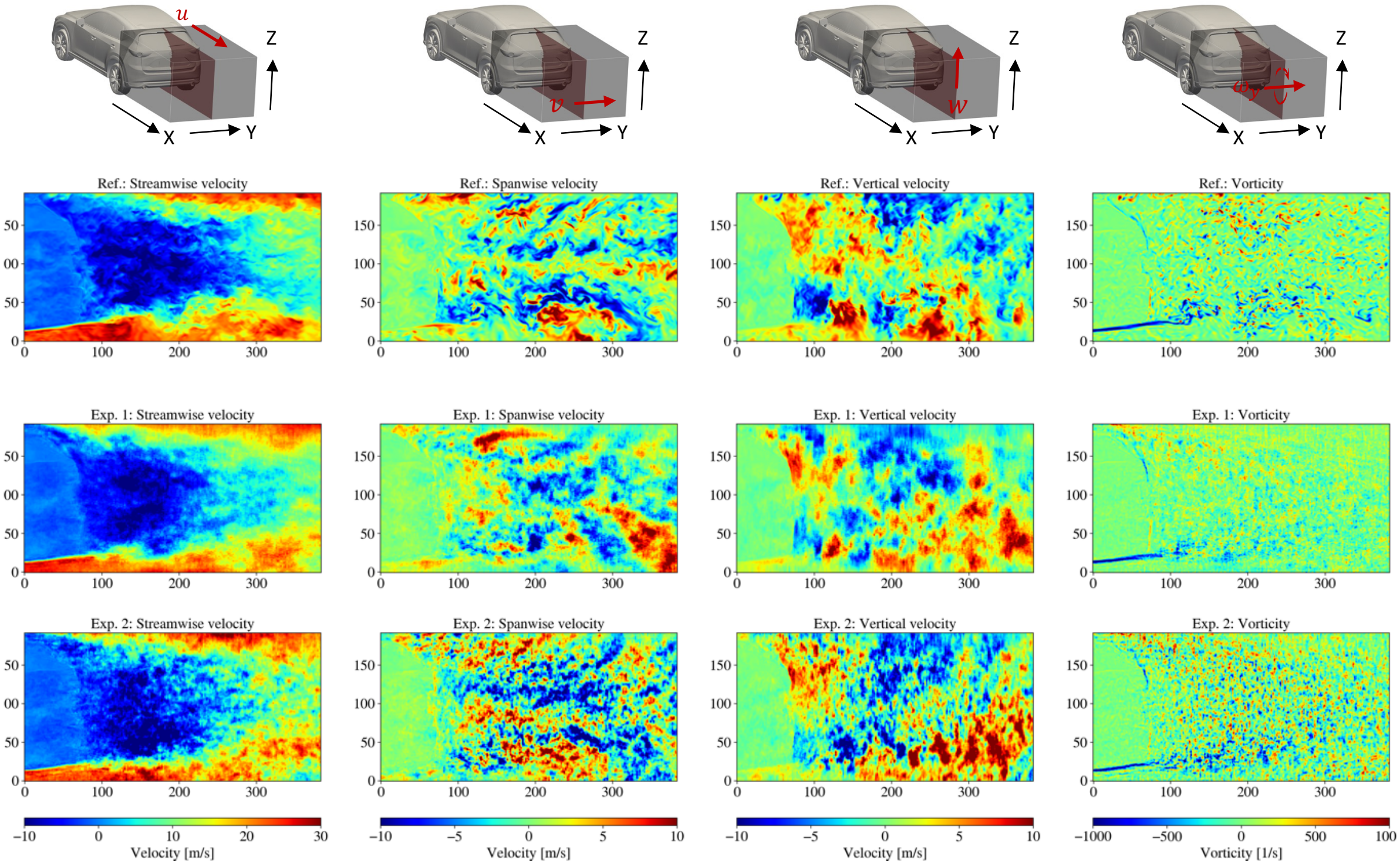

**Figure C.4** Snapshot of the reconstructed flow fields from 32 latent variables for PC1=1.2.

*Appendix D*

Figure D.1 presents a snapshot of the isosurface of the Q-criterion (the second invariant of the velocity gradient tensor).

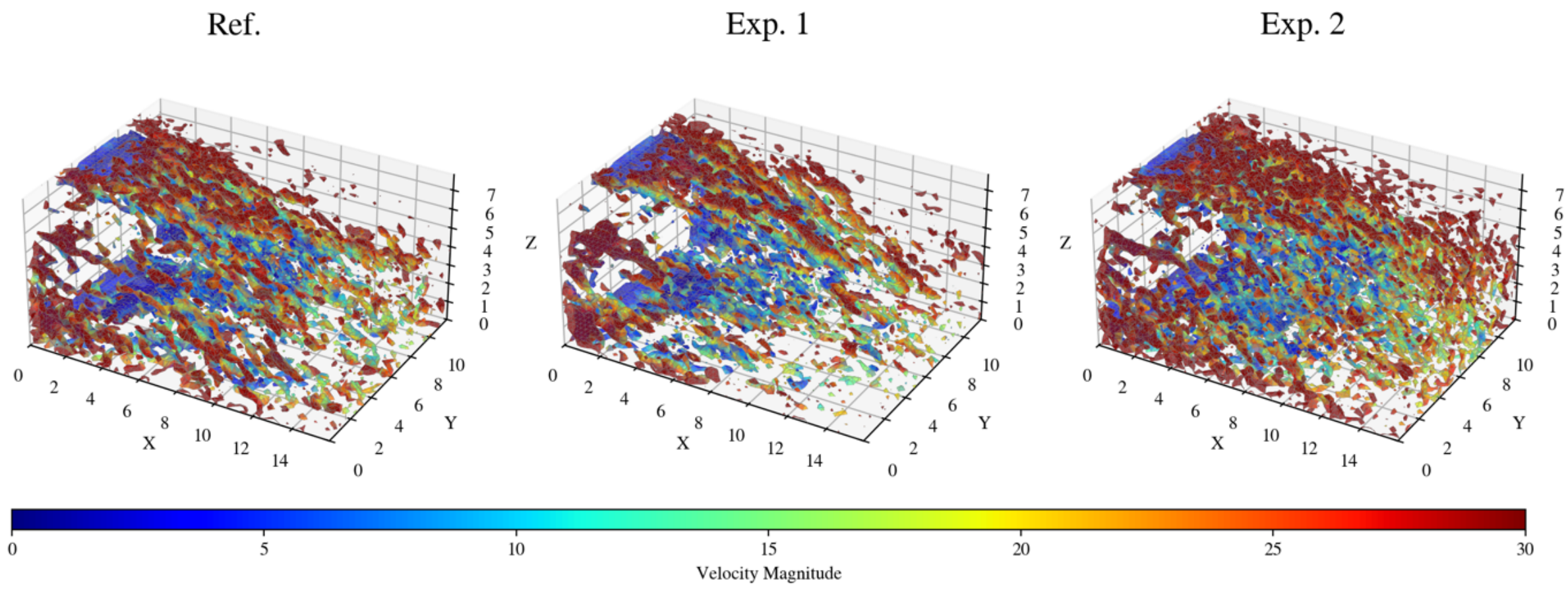

**Figure D.1.** Instantaneous Q-criterion isosurface colored by velocity magnitude for PC1=-1.2.

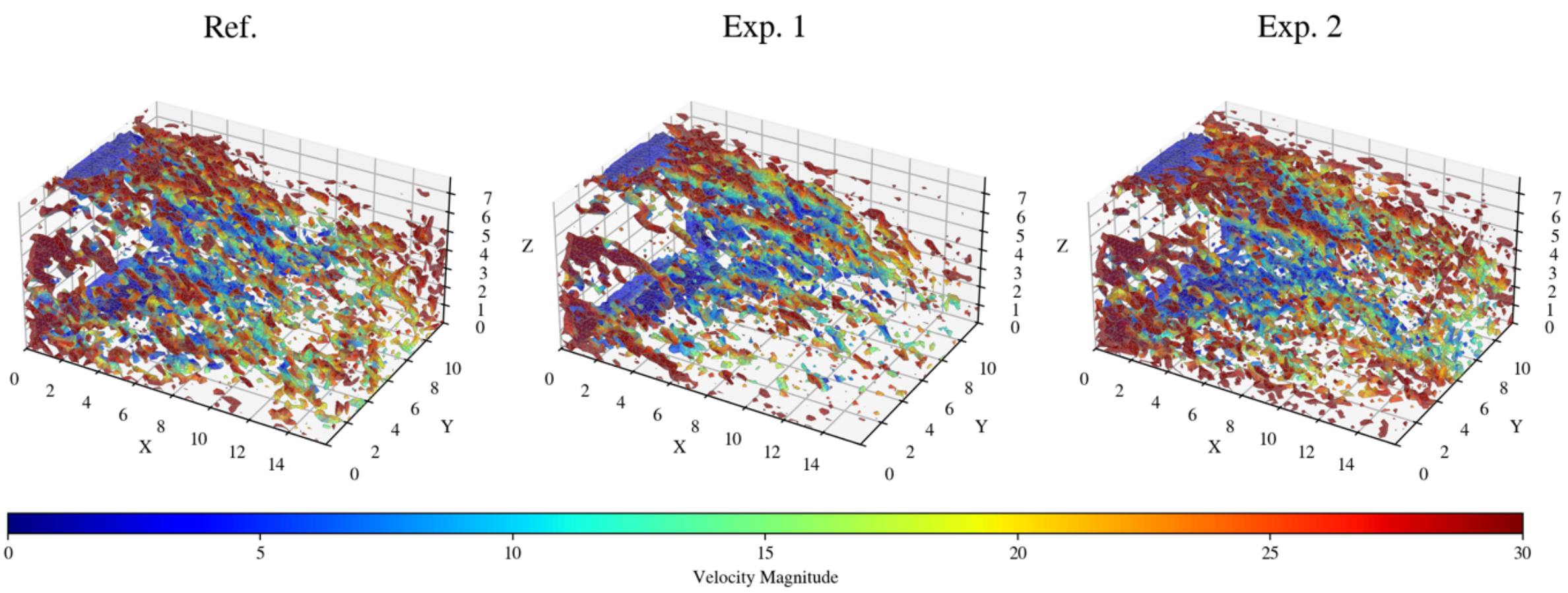

**Figure D.2.** Instantaneous Q-criterion isosurface colored by velocity magnitude for PC1=-0.6.

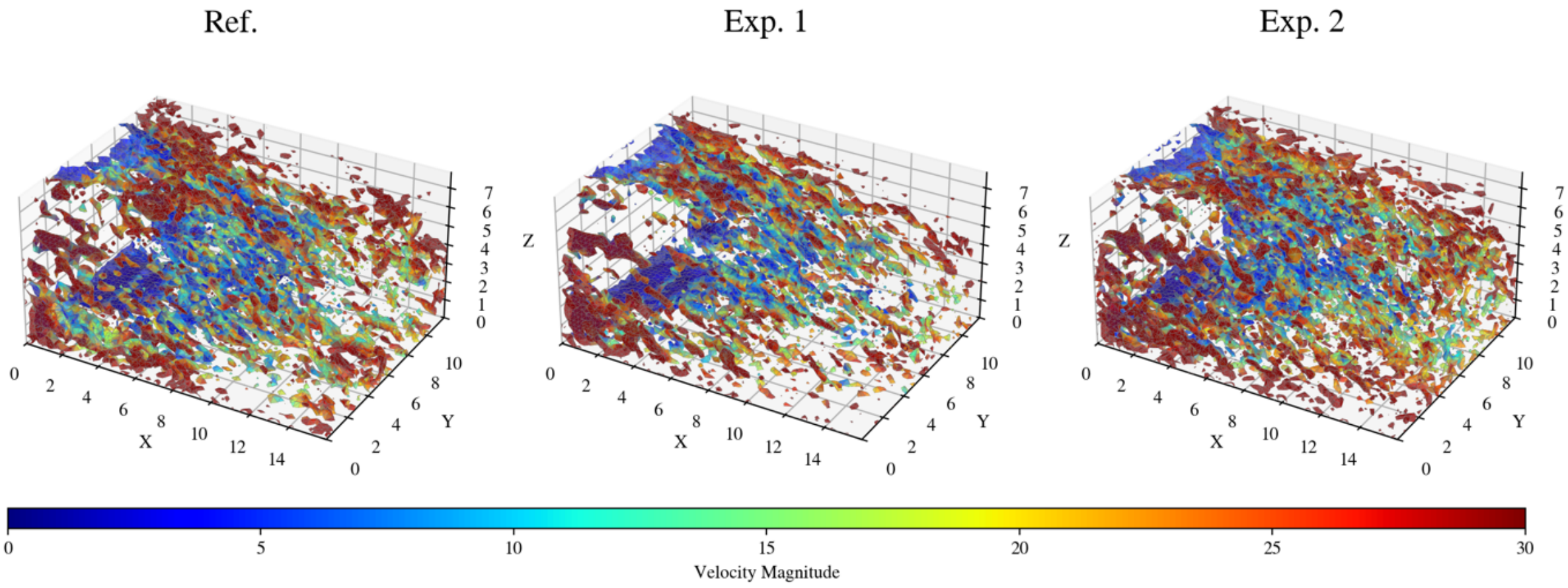

**Figure D.3.** Instantaneous Q-criterion isosurface colored by velocity magnitude for PC1=0.6.

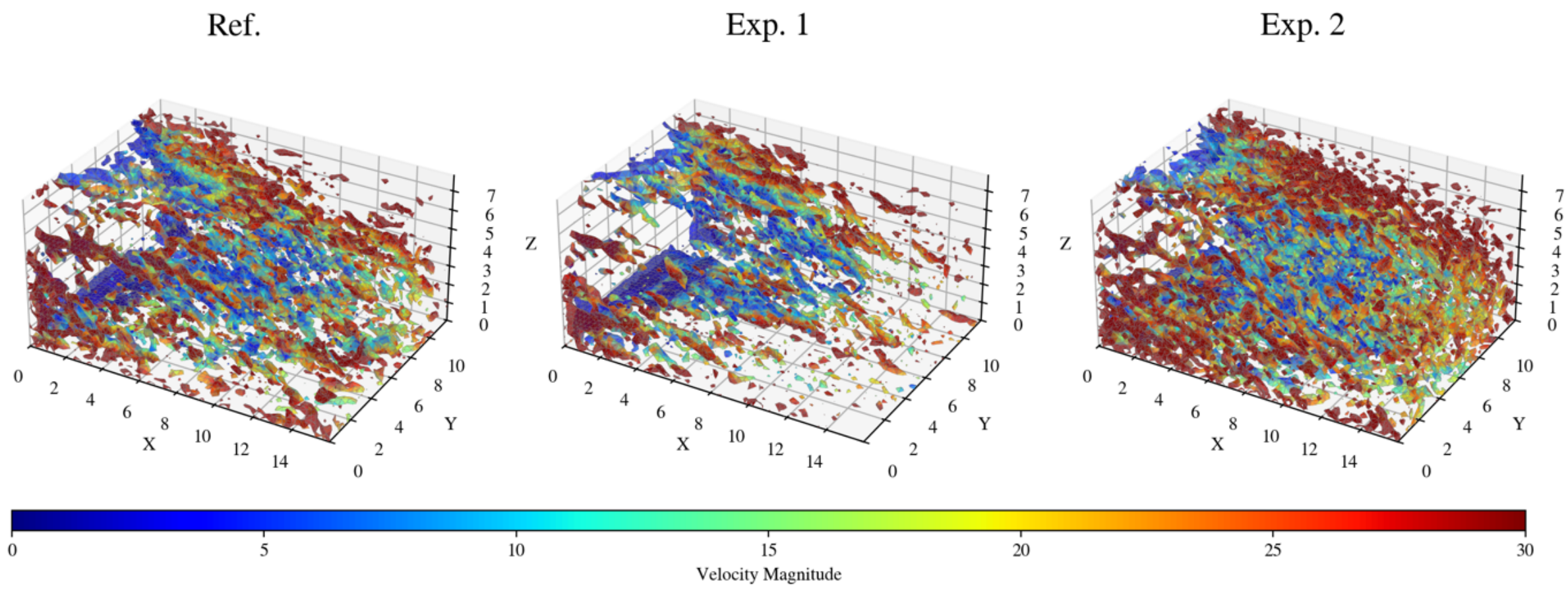

**Figure D.4.** Instantaneous Q-criterion isosurface colored by velocity magnitude for PC1=1.2.

## *Appendix E*

Figure E.1 presents the energy spectrum of the flow field for all vehicle shape patterns.

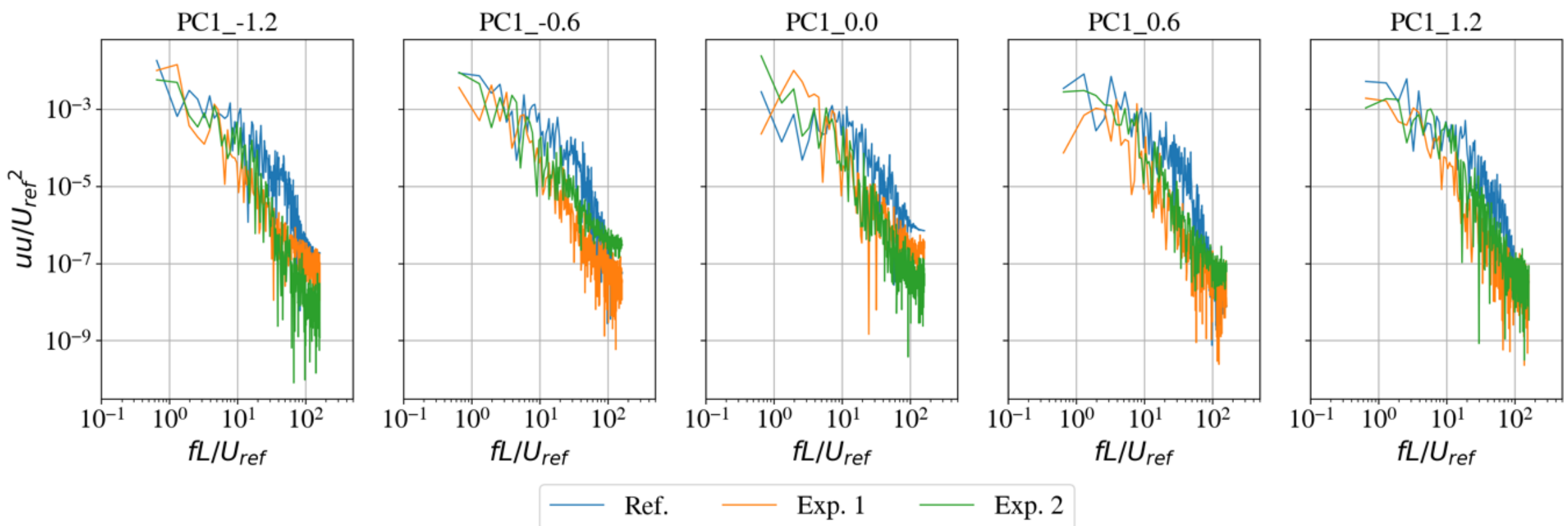

**Figure E.1.** Energy spectra of the flow fields for all vehicle shape variations.

*Appendix F*

Figure F.1 presents the probability density function of vorticity for the flow field of all vehicle shape patterns.

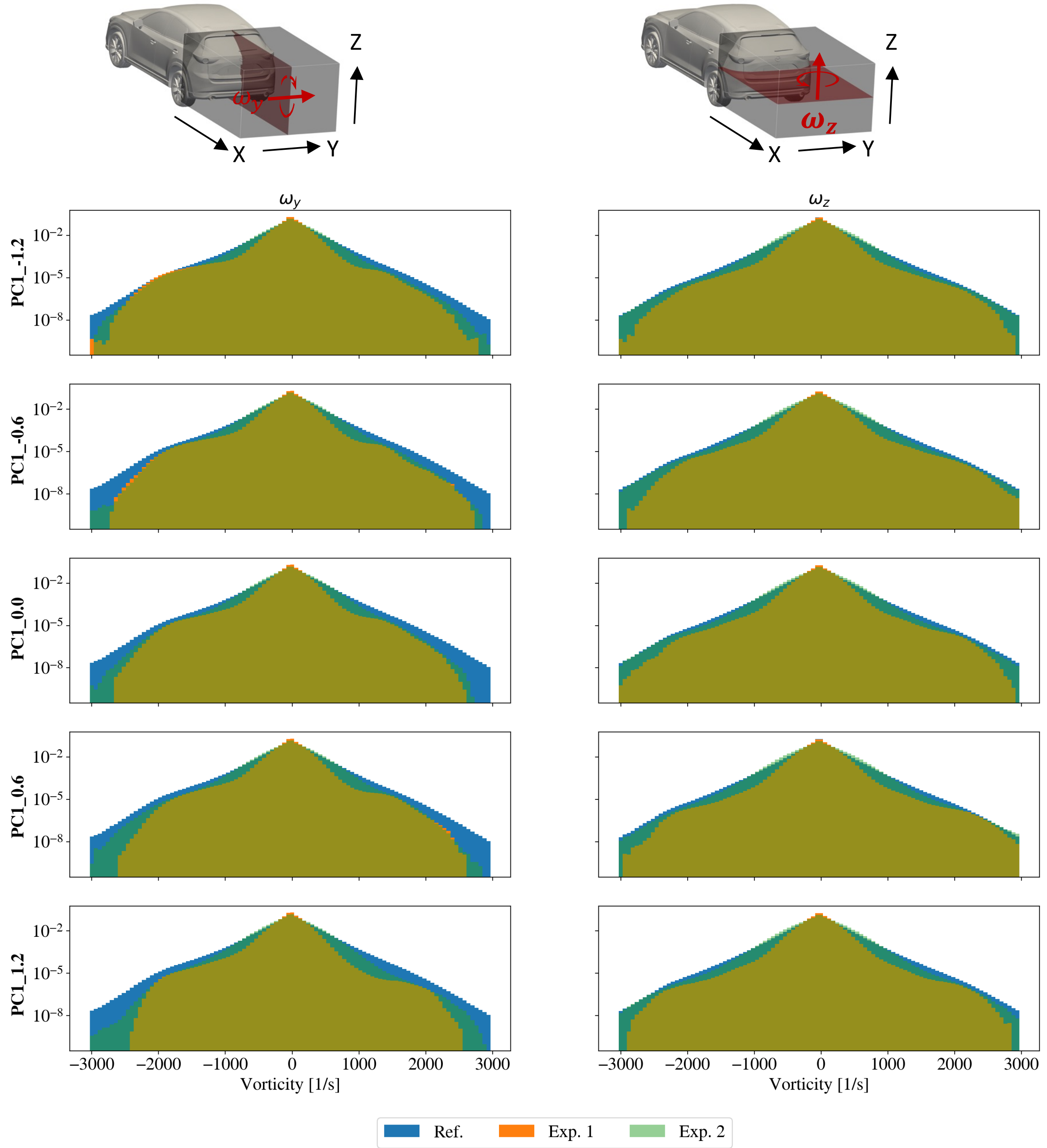

**Figure F.1.** Probability density function of vorticity for the flow fields around all vehicle shape patterns.

*Appendix G*

Figure G.1 presents the time variation of turbulent kinetic energy for the flow field of all vehicle shape patterns.

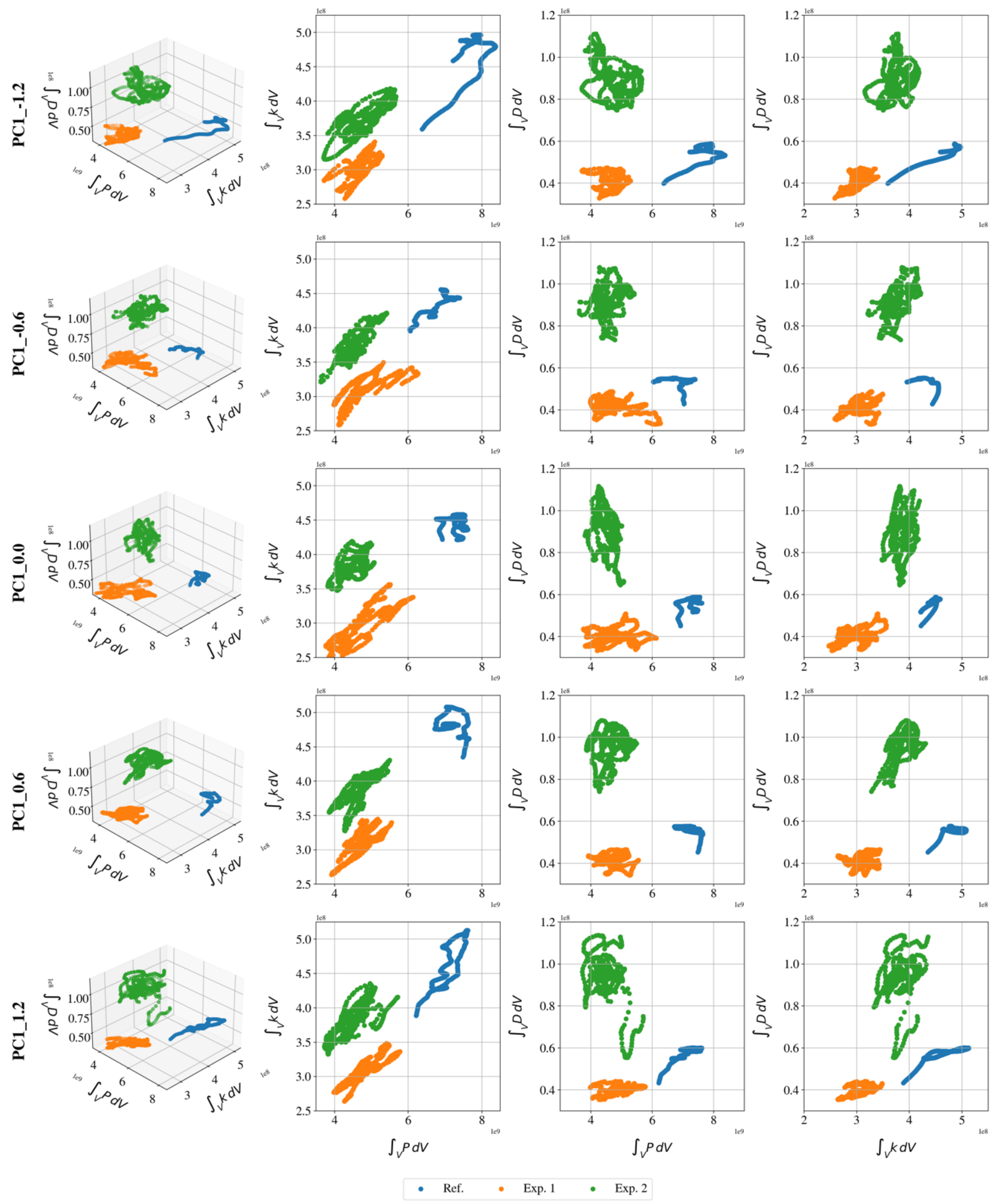

**Figure G.1.** Phase space trajectory of the physical system, defined by turbulent kinetic energy (TKE), TKE production, and TKE dissipation, for all vehicle shape variations.

*Appendix H*

Figure H.1 presents the correlation coefficient of the flow field for all vehicle shape patterns.

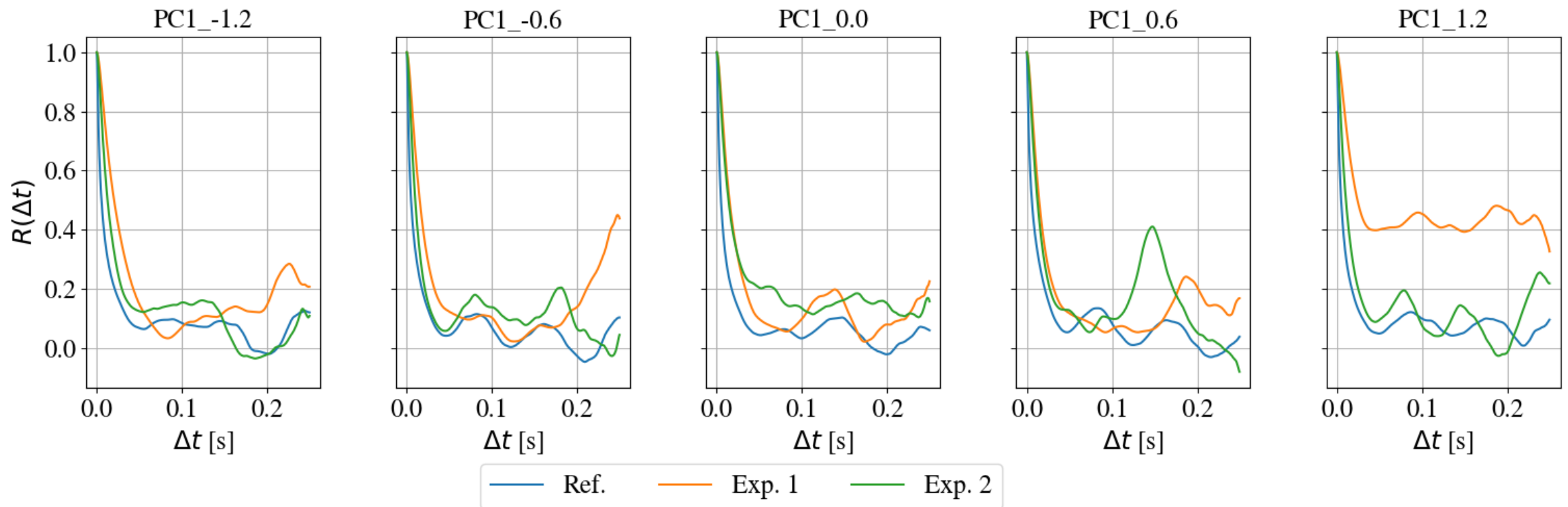

**Figure H.1.** Temporal correlation coefficient for all vehicle shape variations.